\documentclass[twocolumn,showpacs,superscriptaddress,preprintnumbers,amsmath,amssymb,floatfix,10pt]{revtex4}

\newcommand \beq {\begin{equation}}
\newcommand \eeq {\end{equation}}
\newcommand \ben {\begin{eqnarray}}
\newcommand \een {\end{eqnarray}}

\usepackage{wrapfig}
\usepackage{times}
\usepackage[pdftex]{graphicx} 
\usepackage{color}
\usepackage{dcolumn}
\usepackage{bm}
\usepackage{amsmath}
\usepackage{setspace}
\PassOptionsToPackage{caption=false}{subfig}
\usepackage{subfig}
\usepackage{multirow}

\newcommand{\mbfk}{{\mathbf k}}
\newcommand{\mbfq}{{\mathbf q}}
\newcommand{\mbfr}{{\mathbf r}}
\newcommand{\mbfi}{{\mathbf i}}
\newcommand \nline {\nonumber \\}

\begin{document}

\title{A Multi-Component Phase Field Crystal Model for Structural Transformations in Metal Alloys}

\author{Nana Ofori-Opoku} \email{oforion@mcmaster.ca}
\affiliation{Department of Materials Science and Engineering, McMaster University, 1280 Main Street West, Hamilton, Canada L8S-4L7}

\author{Vahid Fallah}
\affiliation{Department of Materials Science and Engineering, McMaster University, 1280 Main Street West, Hamilton, Canada L8S-4L7}
\affiliation{Department of Mechanical and Mechatronics Engineering, University of Waterloo, 200 University Avenue West, Waterloo, Canada N2L-3G1}

\author{Michael Greenwood}
\affiliation{CanmetMaterials, NRCan, 183 Longwood Road South, Hamilton, Canada L8P-0A5}
\affiliation{Department of Materials Science and Engineering, McMaster University, 1280 Main Street West, Hamilton, Canada L8S-4L7}

\author{Shahrzad Esmaeili}
\affiliation{Department of Mechanical and Mechatronics Engineering, University of Waterloo, 200 University Avenue West, Waterloo, Canada N2L-3G1}

\author{Nikolas Provatas}
\affiliation{Department of Physics and Centre for the Physics of Materials, McGill University, 3600 Rue University, Montreal, Canada H3A-2T8}
\affiliation{Department of Materials Science and Engineering, McMaster University, 1280 Main Street West, Hamilton, Canada L8S-4L7}

\begin{abstract}
We present a new phase field crystal model for structural transformations in  multi-component alloys. The formalism builds upon the two-point correlation kernel developed in Greenwood {\it et al.} for describing structural transformations in pure materials [Phys.~Rev.~Lett.~{\bf 105}, 045702 (2010)].  We introduce an effective two-point correlation function for multi-component alloys that uses the local species concentrations to interpolate between different  crystal structures.  A simplified version of the model is derived for the particular case of three-component (ternary) alloys, and its equilibrium properties are demonstrated. Dynamical equations of motion for the density and multiple species concentration fields are derived, and the robustness of the model is illustrated with examples of complex microstructure evolution in dendritic solidification and solid-state precipitation.
\end{abstract}

\pacs{64.70.K-,61.50.Ah,81.10.Aj,46.15.-x}

\maketitle

\section{Introduction}
Engineering alloys require the additions of multiple components to achieve desired properties. This, however, makes the investigation of their microstructure evolution and defect interactions difficult. The properties, and therefore the resultant behaviour, of alloys can directly be correlated to the chemical make up, microstructure and the phase selection processes these alloys undergo upon solidification and subsequent downstream processing, such as thermal treatments. In the case of binary alloys, models of solidification processes such as nucleation, free growth and coarsening kinetics, segregation and second phase formation have been relatively well developed. However, for multi-component alloys, the complex interactions involved between the different chemical species, dislocations and other defects make such phenomena far more difficult to study, even with advances in characterization techniques such as conventional and high resolution transmission electron microscopy.

Advances in modelling have significantly improved our understanding of the fundamental nature of microstructure and phase selection processes. Notable contributions have been made using the phase field methodology (PFM), which has been successful at examining mesoscale microstructure evolution over diffusive time scales. The greatest success of the PFM has come in the area of solidification \cite{Karma98,Provatas98,Echebarria04,Greenwood04,Rappaz03,Boettinger99,Granasy03}. The phase field concept has gone far beyond its origins. It is now capable of describing, through the introduction of various auxiliary fields, a wealth of phenomena such as multiple crystal orientations \cite{Warran00,Warren03,Granasy04a}, multiple components and phases \cite{Steinbach06,Steinbach09,Nestler12}, defect-solute interactions \cite{Haataja04}, elasticity \cite{Zhu04a,Fan96} and plasticity \cite{Wang01}.

There has recently emerged an atomic-scale modelling formalism, called the phase field crystal model (PFC)~\cite{Elder02,Elder07}. This method, operates on atomistic length scales and diffusive time scales and self-consistently incorporates elasticity,  multiple crystal orientations, grain boundaries, dislocations, and the evolution of microstructure on diffusive time scales. For both pure materials and binary alloys, Elder and co-workers~\cite{Elder07} and Jin and Khachaturyan~\cite{Jin06}  have shown that PFC models can be formally derived from classical density functional theory (CDFT), where the order parameter can be related to the atomic probability density \cite{Akusti09}. As such, many basic microstructure phenomena can be seen as arising self-consistently  from a simple fundamental theory described by a small set of physically motivated  parameters. With the ability of the PFC density field to also assume disordered states, it is also possible to examine amorphous or glassy states ~\cite{Berry08,Archer12}. Phase field crystal models are also exceedingly simple to work with numerically. The use of coarse graining approaches has further shown that PFC-type models can be used as generators of traditional phase field models, as well as so-called {\it amplitude} models, essentially phase field models with complex order parameters. These make it possible to simulate different crystal orientations and defect structures on mesoscopic length and time scales~\cite{Huang08,Wu07,Majaniemi09,Provatas10}, and also exploit the scaling afforded by adaptive mesh refinement ~\cite{Athreya07}.

A weakness of the early PFC models was their inability to systematically describe and control complex crystal structures and coexistence between them. Greenwood {\it et al.}~\cite{Greenwood10,Greenwood11} addressed this shortcoming  by introducing a class of multi- peaked, two-point direct correlation functions that contained some of the salient features of CDFT, but retained the simplifications that gave the original PFC formalism its numerical efficiency. This so-called {\it ``XPFC''} formalism was later extended to binary alloys, and applied to phenomena such as eutectic solidification and elastic anisotropy ~\cite{GreenwoodOfori11}, solute drag \cite{Greenwood12}, quasi-crystal formation \cite{rottler12}, solute clustering and precipitation mechanisms in simplified Al-Cu alloys \cite{Fallah12a} and 3D stacking fault structures in FCC crystals \cite{Berry12}.

In this paper, we generalize the XPFC formalism of Greenwood {\it et. al} to the case of $N$-component alloys. The approach begins with the truncated CDFT energy functional of an $N$-component system. At the core of our excess free energy are the particle interactions of Ref.~\cite{Greenwood10,Greenwood11}, adapted for different structural phases in alloys by making the interaction kernel a function of the local species concentrations. We compute the equilibrium properties of our model for the case of a ternary alloy and compare the resulting model phase diagram to an experimental ternary system. The dynamics of the model are then demonstrated in the context of dendritic solidification and solid-state precipitation

The remainder of this paper is organized as follows. We begin with deriving the full $N$-component XPFC energy functional in Section~\ref{PFC-multi-functional} from a simplified, truncated classical density function theory of freezing similar to that of Ramakrishan and Yussouff~\cite{Ramakrishan79}. We then derive a second, simplified version of the model that is the $N$-component analogue of the previous binary XPFC model in literature. Section~\ref{ternary}  calculates the equilibrium properties of the model for a particular case of a ternary system via isothermal sections of the phase diagram.  Section~\ref{ternary-apps} presents some  numerical examples of microstructure evolution by simulating dendritic solidification and solid-state precipitation. We end with a summary and conclusions.

\section{XPFC Energy Functionals for $N$-Component Alloys}
\label{PFC-multi-functional}
A general free energy functional for an $N$-component alloy is derived starting from the classical density functional theory of freezing energy formalism of Ramakrishan and Yussouff~\cite{Ramakrishan79}, where each alloy component is written in terms of a density field $\rho_i$. The model is re-written in terms of total density and concentration variables to make contact with standard models used in the description of alloys.  The model is then collapsed to a simplified form of the free energy, similar to the simplified form for the binary XPFC model of Greenwood {\it et al.}~\cite{GreenwoodOfori11}. Finally, equations of motion for the total density and each concentration field are presented for both versions of the model free energy.

\subsection{Deriving an XPFC Energy Functional for $N$-component Systems}
The free energy functional of an $N$-component mixture can be described by two contributions; a local free energy for each of the $N$ density fields and an excess free energy due to species interactions. The local free energy is treated as an ideal energy which drives the density fields to become uniform. The excess contribution drives the density fields to become periodic by creating minima in the free energy for these states. We can write the free energy functional of the mixture as
\beq
\frac{\Delta {\mathcal F}}{k_B\, T}=\int d\mbfr~\left\{\frac{\Delta F_{id}}{k_B\, T}+\frac{\Delta F_{ex}}{k_B\, T}\right\},
\label{truncated-DFT-energy}
\eeq
where $\Delta F_{id}$ denotes the ideal energy and $\Delta F_{ex}$ is the excess energy which accounts for interactions between atoms through correlative interactions. This latter term, gives rise to structural symmetry, elasticity and interactions between topological defects. The constant $k_B$ is the Boltzmann constant and  $T$ the temperature. The differential $d\mbfr \equiv dx dy dz$.

The ideal energy, $\Delta F_{id}$, gives the entropic contribution for an $N$-component system. For small density changes from a reference density of each component, it is defined as
\beq
\frac{\Delta F_{id}}{k_BT } =  \sum_i^N \rho_{i} \ln \left(\frac{\rho_{i}}{\rho_{i}^{o}}\right)-\delta\rho_{i},
\label{idealenergy-log}
\eeq
where $N$ denotes the number of components, which are denoted as $A,B,C$,\ldots,etc., $\rho_i$ is the density of component $i$, and ${\rho_{i}^{o}}$ is the reference density of component $i$ in the liquid phase at co-existence. Following previous PFC models~\cite{Elder07}, we define a total mass density $\rho = \sum_i^N \rho_{i}$ and the total reference mass density as $\rho^o = \sum_i^N \rho_{i}^o$. Following Refs.~\cite{Elder07,Provatas10,GreenwoodOfori11}, we  define concentrations as $c_{i} = \rho_{i}/\rho$ and the corresponding reference compositions by $c_{i}^o = \rho_{i}^o/\rho^o$. Furthermore, for convenience we define a dimensionless mass density of the form $n = \rho/\rho^o-1$. With these definitions and the conservation condition $\sum_i c_i \equiv 1$, Eq.~(\ref{idealenergy-log}) simplifies to the dimensionless form
\beq
\frac{\Delta F_{id}}{k_{B}T\rho^o} \!=\! (n\!+\!1) \ln \left(n+1\right) \!-\! n \!+\!
\left(n\!+\!1\right)\sum_i^N c_i\ln{\frac{c_{i}}{c_{i}^{o}}}.
\label{dimlss-ideal-energy}
\eeq

The excess energy takes into account inter-particle interactions truncated at two-particle, i.e., $A$-$A$,$B$-$B$,\ldots, $N$-$N$, $A$-$B$,\ldots,$A$-$N$, $\cdots$ interactions. This can be defined as,
\begin{align}
\frac{\Delta F_{ex}}{k_B\,T} &= \!-\frac{1}{2}\! \int d\mbfr^{\prime}\,\sum_i^N\sum_j^N\delta\rho_i\left(\mbfr\right)\,_2^{ij}\left(\mbfr,\mbfr^\prime\right)\,
\delta\rho_j\left(\mbfr^\prime\right),
\label{two-particle-interaction}
\end{align}
where $C_2^{ij}$ represent all combinations of  two-particle correlations, in this work assumed isotropic (i.e., $C_2^{ij}\left(\mbfr,\mbfr^\prime\right)=C_2^{ij}\left(|\mbfr-\mbfr^\prime|\right)$), between the field describing species $i$ and $j$, respectively, where $i,j=A,B,C,\ldots,N$. We write Eq.~(\ref{two-particle-interaction}) in terms of the reduced density $n$ and compositions $c_i$. As in Refs.~\cite{Elder07,GreenwoodOfori11}, we consider only the lowest order contributions of the compositions $c_i$, which vary on length scales much larger than the density $n$, which are periodic on the scale of the lattice constant. This allows us to simplify integrals arising from Eq.~(\ref{two-particle-interaction}), which couple $c_i(\mbfr^\prime)$ together with $n(\mbfr^\prime)$
\footnote{This approximation captures the separation of scales between concentration and density adequately in direct PFC
simulations. However, it is not suitable when coarse graining the model to derive corresponding of complex amplitude equations.
In the latter case, slow fields must be expanded to second order Taylor series in powers of ${(\mbfr - \mbfr^\prime)}$.}.
For example,
\beq
\int d\mbfr^\prime \, C_2^{ij}(|\mbfr-\mbfr^\prime|) n(\mbfr^\prime) c_i(\mbfr^\prime) \approx c_i(\mbfr) \int d\mbfr^\prime C_2^{ij}(|\mbfr-\mbfr^\prime|) n(\mbfr^\prime).
\nonumber
\eeq
To simplify notation, the notation $n(\mbfr^\prime) \equiv n'$ and $c_{i}(\mbfr^\prime) \equiv c_{i}'$ is used hereafter. With these simplifications and notations, the excess energy of Eq.(\ref{two-particle-interaction}) can be written in terms of the dimensionless variables $n$ and $\{c_i\}$ as
\begin{align}
\label{dimlss-excess-energy}
\frac{\Delta F_{ex}}{k_BT\rho^o} &=-\frac{1}{2}\sum_{i,j}^N \int d\mbfr\left[n\,c_{i}\,c_{j}\, + c_{i}\,c_{j} - c_{i}^{o}\,c_{j}\right]\,\int d\mbfr^{\prime}\, C_2^{ij}n^\prime\nonumber \\
&-\frac{1}{2}\sum_{i,j}^N \int d\mbfr\left[n\,c_{i} + c_{i} - c_{i}^{o}\right]\,\int d\mbfr^{\prime}\, C_2^{ij}c_{j}^{\prime}\\
&-\frac{1}{2}\sum_{i,j}^N \int d\mbfr\left[c_{i}^{o}\,c_{j}^{o} - n\,c_j^o \,c_{i} - c_j^o \, c_{i} \right]\,\hat{C}_2^{ij}(|\mbfk| \!=\!0),\nonumber
\end{align}
where $\hat{C}_2^{ij}$ is the Fourier transform of $C_2^{ij}(|\mbfr -\mbfr^{\prime}|)$, and satisfies
\beq
\hat{C}_2^{ij}(|\mbfk| \!=\!0) = \int d\mbfr^{\prime}\, C_2^{ij}(|\mbfr -\mbfr^{\prime}|),
\label{kzero}
\eeq
and where we have introduced the notation $C_2^{ij}\equiv \rho^{o}C_2^{ij}\left(|\mbfr-\mbfr^\prime|\right)$, which is the direct two-point correlation function.

Collecting terms from Eqs.~(\ref{dimlss-ideal-energy}) and (\ref{dimlss-excess-energy}) gives the complete $N$-component free energy functional, written in dimensionless form,
\begin{align}
\label{multi-full-energy}
\frac{\Delta {\mathcal F}}{k_B\,T\rho^{o}} &=
\int d\mbfr~(n+1) \ln \left(n+1\right) \! - \! n \! + \!  \Delta F_{\text {mix}} \left( \{c_i\} \right)(n \! +\! 1) \nonumber \\
&-\frac{1}{2}\sum_{i,j}^N \int d\mbfr\left[n\,c_{i}\,c_{j}\, + c_{i}\,c_{j} - c_{i}^{o}\,c_{j}\right]\,\int d\mbfr^{\prime}\, C_2^{ij}n^\prime\nonumber \\
&-\frac{1}{2}\sum_{i,j}^N \int d\mbfr\left[n\,c_{i} + c_{i} - c_{i}^{o}\right]\,\int d\mbfr^{\prime}\, C_2^{ij}c_{j}^{\prime} \\
&-\frac{1}{2}\sum_{i,j}^N \int d\mbfr\left[c_{i}^{o}\,c_{j}^{o} - n\,c_{i}\,c_{j}^{o} - c_{j}^{o} \, c_{i}\right]\,\hat{C}_2^{ij}(|\mbfk|\!=\!0),
\nonumber
\end{align}
where $\Delta F_{\text {mix}}(\{c_i\})$ denotes the ideal entropy of mixing,
\beq
\Delta F_{\text {mix}}(\{c_i\}) = \sum_i^N c_i\ln{\frac{c_{i}}{c_{i}^{o}}}.
\label{mixing_entropy}
\eeq
Equation~(\ref{multi-full-energy}) is the full $N$-component PFC model in CDFT form. When a form for $C_2^{ij}$ is specified,  it can be used directly. However, this form is not convenient to make contact with other theories and models in the literature. It will be transformed into a simpler form in the next section.

\subsection{Simplified $N$-Component XPFC Free Energy}
\label{simplified-PFC-multi-functional}
It is instructive to reduce the model of Eq.~(\ref{multi-full-energy}) to a minimal form that retains the salient features of the original model but can also make contact with previous PFC and phase field models. To do so, certain simplifications must be made.

First, an expansion of the ideal free energy term is taken to fourth order in the limit of small $n$, i.e., around the reference $\rho^{o}$. The logarithms in the entropy of mixing (Eq.~(\ref{mixing_entropy})) are left unexpanded for convenience. Secondly, the terms with correlation kernels can be simplified by retaining the long wavelength behaviour of all compositions $c_i$, where they vary much more slowly than $n$. Following the procedures outlined in Refs.~\cite{Majaniemi09,Provatas10,Elder10,Huang10}, it can be shown that upon coarse graining, all terms containing linear powers of $n$ or $n'$ in
Eq.~(\ref{multi-full-energy}) vanish. Also, terms containing only concentration fields and a correlation function give rise to local products of $c_i\, c_j$ (which arise from the $\mbfk=0$ part of $C_2^{ij}$, and look analogous to the last term in Eq.~(\ref{multi-full-energy})) and products between their corresponding gradients. The reader is referred to Appendix~\ref{long-wavelength-limit} for details of the coarse graining procedure applied to terms of Eq.~(\ref{multi-full-energy}).  After some tedious but straightforward algebra, the above approximations lead to the following simplified $N$-component
XPFC free energy functional,
\begin{align}
{\cal{F}} &= \int d\mbfr~\Bigg\{\frac{n^2}{2}
\!-\!\eta \frac{n^3}{6} \!+\! \chi \frac{n^4}{12} \!+\! \omega \, \Delta F_{\text {mix}}(\{c_i\}) (n+1) \nonumber \\
&\!-\! \frac{1}{2}n \!\! \int d\mbfr^\prime  C_{eff}(|\mbfr-\mbfr^\prime|)\,n^\prime \!+\! \frac{1}{2}\sum_{i,j}^{N} \kappa_{ij} \nabla c_i \cdot \nabla c_j \Bigg\},
\label{simplified-Energy}
\end{align}
where
\beq
C_{eff}(|\mbfr-\mbfr^\prime|) = \sum_{i,j=1}^N c_i\,c_j \,C_2^{ij}(|\mbfr-\mbfr^\prime|).
\label{interpol1}
\eeq
The parameters $\eta$, $\chi$ and $\omega$ are constants, the significance of which is discussed further below. The $\kappa_{ij}$ are gradient energy
coefficients associated with compositional interfaces involving $c_i$ and $c_j$. For notational convenience,  ${\cal F}$ is used to denote $\Delta { \cal F }/ k_B T \rho^o$.

The  parameters $\eta$ and $\chi$ corresponding to Eq.~(\ref{multi-full-energy}) are formally equal to one, but hereafter will be treated as free
parameters that can be used to correct the density dependence of the ideal free energy away from the reference density $\rho^o$, i.e., to match the bulk free energy to materials properties. Also, it was shown in Ref.~\cite{Huang10} that the $\mbfk=0$ mode of higher-order correlation terms in a CDFT expansion will contribute local polynomial terms in $c_i$ and $n$, analogous to the  $\hat{C}_2^{ij}(|\mbfk|=0)$ terms of  Eq.~(\ref{multi-full-energy}).  These terms can be combined with an expansion of the $\Delta F_{\text{mix}}$  term in Eq.~(\ref{multi-full-energy}) to produce a messy polynomial expansion of the local free energy in powers of the elements of $\{c_i\}$ and $n$.  To keep the form of  the free energy compact, we have found that it is simpler to introduce a parameter, $\omega$, which modifies the mixing free energy from its ideal form, away from  the reference  compositions $c_i^o$. Finally, in the present work, the gradient energy coefficient tensor will be assumed to be diagonal for simplicity, i.e., $\kappa_{ij}=0$ for $i \ne j$ and $\kappa_{ii} >0$ for all $i$.

The correlation function in Eq.~(\ref{interpol1}) is too basic to capture the properties of very complex alloys --although it can capture some properties of simple alloys. Guided by the form of the first term on the second line of Eq.~(\ref{multi-full-energy}), it can be seen that higher-order correlation functions will contribute terms of the form $c_i \,c_j\, c_k\,C_3^{ijk}$, $c_i\,c_j\,c_k \,c_l C_4^{ijkl}$, etc. To emulate such higher-order non-local contributions effectively, we introduce an effective correlation function of the form
\beq
C_{eff}(|\mbfr-\mbfr^\prime|) = \sum_{i=1}^N X_i(\{c_j\})\,C_2^{ii}(|\mbfr-\mbfr^\prime|).
\label{ceff}
\eeq
The $X_i$ are as yet undetermined polynomial functions of the elements of $\{c_j\}$. The role of the $X_i$ is to determine the resultant local crystalline
structure by interpolating between the kernels $\hat{C}_2^{ii}$ (defined below), which define the base equilibrium crystal structures of each pure component $i$.
The interpolation is done through appropriately constructed polynomial expansions of the elements of $\{c_j\}$.  The order of  $X_i$ depends on the number of components in the system and can be made as high as required to smoothly interpolate from one correlation kernel to another. We have found that Eq.~(\ref{ceff}), through appropriate choices of $X_i$, combined with other model parameters, is robust enough to model a wide variety of alloy systems.

The model in Eq.~(\ref{simplified-Energy})  captures the usual features of other PFC models, while allowing for a very easy control of a wide range of crystal structures in different phases. It is motivated from considerations of classical density functional theory but simplified enough to make numerically tractable simulations possible, as will be shown below. Finally, we note that the form of the  expansion in Eq.~(\ref{ceff}) is dimensionally motivated from higher-order terms in CDFT but is flexible enough to model experimentally relevant multi-component alloys quantitatively using, for example, thermodynamic databases.

\subsection{Dynamics}
\label{dynamics}
Equations of motion for the density $n$ and each of the concentration fields $c_i$ follow conserved dissipative dynamics. Namely the dimensionless
density $n$ obeys
\begin{align}
\label{dynamics-n}
\frac{\partial n}{\partial t} \!\! &= \!\nabla \! \cdot \! \left( M_n\nabla \frac{\delta {\cal{F}}}{\delta\,n} \right) +\zeta_n \\
\!\! &=\!\! \nabla \! \cdot \! \left( \! M_n \! \nabla\Biggl\{n-\eta \frac{n^2}{2} + \chi \frac{n^3}{3} \!+\! \omega\Delta F_{\text {mix}}(\{c_i\}) \!-\! C_{eff}\,n \! \Biggr\} \! \right)  \nonumber \\
&+\zeta_n,\nonumber
\end{align}
while the dynamics of each composition field, $c_i$, evolve according to
\begin{align}
\label{dynamics-c}
\frac{\partial c_i}{\partial t} \!\! &=\! \nabla \! \cdot \! \left( M_{c_i} \nabla \frac{\delta {\cal F}}{\delta c_i} \right) + \zeta_{c_i} \\
\!\! &=\!\!\nabla\! \cdot \!\left(\! M_{{c}_{i}} \!\nabla\Biggl\{\omega(n+1)\frac{\delta \Delta F_{\text {mix}}}{\delta c_i}
\!-\!\frac{1}{2}n\frac{\delta C_{eff}}{\delta c_i}\,n \!-\! \kappa_i\nabla^2c_i \!\Biggr\} \!\right) \nonumber \\
&+ \zeta_{c_i},\nonumber
\end{align}
where the following shorthand notations have been made,
\begin{align}
C_{eff} \, n &\equiv \int d {\bf r}' C_{eff} ({\bf |r-r'|}) n({\bf r}') \nonumber \\
n \, \frac{\delta C_{eff} }{\delta c_i} \, n  &\equiv n({\bf r})  \int d {\bf r}' \frac{\delta  C_{eff}}{\delta c_i} ({\bf |r-r'|}) n({\bf r}').
\label{shothand}
\end{align}
The coefficients $M_n$ and $M_{c_{i}}$ denote the mobility of the density and each concentration, respectively, and strictly speaking can be functions of the fields. The noise terms $\zeta_n$ and $\zeta_{c_i}$
model coarse grained thermal fluctuations on density and concentrations $c_i$, respectively. They formally satisfy $\langle \zeta_q ({\bf r},t) \zeta_q ({\bf r}',t')
\rangle =- A \nabla^2 \chi_a({\bf r-r'}) \delta (t-t') $, where $q$ denotes the density or one of the concentration fields, $A \propto M_q k_B T $ and $\chi_a({\bf r-r'})$
is the inverse Fourier transform of a Gaussian function, which, following Tegze and co-workers Ref.~ \cite{Tegze11}, can be generalized to have a high frequency
cut off for frequencies above $2 \pi/a$, where $a$ is the lattice constant.  The precise form of $A$, which sets the scale of the thermal fluctuations is not
properly understood in the context of PFC modelling but is the object of several investigations. In the applications illustrated in this paper, the noise is left out of
simulations.

\section{Ternary Systems}
\label{ternary}
In this section, we reduce the simplified free energy functional of section~(\ref{simplified-PFC-multi-functional}) to the case of
three-components, or ternary alloys. We first describe the ternary free energy functional, followed by a discussion of the effective correlation function chosen for ternary systems. With the free energy and effective correlation in hand, we  demonstrate the equilibrium properties of our model by calculating the ternary phase diagrams for a generic $A$-$B$-$C$ system and a simplified Al-Cu-Mg system.

\subsection{Free Energy Functional}
Specializing Eq.~(\ref{simplified-Energy}) for 3-components, denoted here as
$A$, $B$ and $C$, reduces it to
\begin{align}
{\cal F}^{\text{ter}} &= \int d\mbfr~\Bigg\{\frac{n^2}{2} -\eta \frac{n^3}{6} + \chi \frac{n^4}{12} + \omega \Delta F_{\text {mix}}^{\text{ter}} \left( n+1\right)
\label{3-simplified-Energy} \\
&-\frac{1}{2}n\int d\mbfr^\prime C_{eff}^{\text{ter}}(|\mbfr-\mbfr^\prime|)\,n^\prime + \frac{\kappa_{A}}{2}|\nabla c_{A}|^2 + \frac{\kappa_{B}}{2}|\nabla c_{B}|^2 \Bigg\},
\nonumber
\end{align}
where
\begin{align}
\Delta F_{\text {mix}}^{\text{ter}} &= c_{A}\ln{\frac{c_{A}}{c_{A}^{o}}}
+ c_{B}\ln{\frac{c_{B}}{c_{B}^o}}\nline &+(1-c_{A}-c_{B})\ln{\frac{(1-c_{A}-c_{B})}{1-c_{A}^o-c_{B}^o}},
\label{entropy-mixing}
\end{align}
and
\begin{align}
\label{CorrEff}
&C_{eff}^{\text{ter}}(|\mbfr-\mbfr^\prime|) =  X_A(c_{A},c_{B}) C_2^{AA}(|\mbfr-\mbfr^\prime|)\\
 &+  X_B(c_{A},c_{B})\,C_2^{BB}(|\mbfr-\mbfr^\prime|) +  X_C(c_{A},c_{B})\,C_2^{CC}(|\mbfr-\mbfr^\prime|). \nonumber
\end{align}
In arriving at Eq.~(\ref{3-simplified-Energy}), the conditions $c_C = 1-c_A-c_B$ and $c_C^o = 1-c_A^o-c_B^o$  have been used, and the cross gradient  concentration terms in $A$ and $B$ have been neglected.

The effective ternary correlation kernel, $C_{eff}^{\text{ter}}$, is defined by $X_i$ such that $X_A+X_B+X_C \equiv 1$ at all compositions, analogous to the case for the XPFC binary model~\cite{GreenwoodOfori11}. Their particular form, is chosen here to model the  generic properties of eutectic systems. However, by careful alteration of other parameters, other alloy systems can be modelled, e.g. isomorphous and peritectic systems~\cite{GreenwoodOfori11}. Here the $X_i$ used are,
\begin{align}
\label{interp-func}
X_A(c_A,c_B) &= 3c_A^2+2c_Ac_B-2c_A^3-2c_A^2c_B-2c_Ac_B^2\nonumber \\
X_B(c_A,c_B) &= 2c_Ac_B+3c_B^2-2c_A^2c_B-2c_Ac_B^2-2c_B^3\nonumber \\
X_C(c_A,c_B) &= 1-3c_A^2+2c_A^3-3c_B^2+2c_B^3-4c_Ac_B\nonumber \\
&+4c_A^2c_B+4c_Ac_B^2.
\end{align}

\subsection{Correlation Functions $C_2^{ii}$}
The XPFC formalism is best suited for numerical simulation in Fourier space. The pure component correlation functions $C_2^{ii}(|\mbfr -\mbfr'|)$ are thus constructed directly in Fourier space, where they are denoted $\hat{C}^{ii}_2(\mbfk)$.  Each  component, $i$, contributes a correlation function that supports the desired equilibrium crystal structure for a pure component. A Fourier space peak of $\hat{C}^{ii}_2(\mbfk)$~\cite{Greenwood11}, for a given mode, $j$,
is denoted by
\beq
\hat{C}^{ii}_{2j}=e^{-\frac{\sigma^2}{\sigma^2_{Mj}}}e^{-\frac{(k-k_j)^2}{2\alpha^2_j}}.
\label{CorrF}
\eeq
The total correlation function for component $i$, $\hat{C}_2^{ii}$, is  defined by the envelope of all peaks $\hat{C}^{ii}_{2j}$. The first exponential in Eq.~(\ref{CorrF}) sets the temperature scale via a Debye-Waller prefactor that employs an effective temperature parameter, $\sigma$. We also define an effective transition temperature, $\sigma_{Mj}$, which subsumes the effect of planar and atomic densities associated with the family of planes corresponding to mode
$j$ ~\cite{GreenwoodOfori11}. The second exponential sets the position of the reciprocal space peak at $k_j$, which defines the inverse of the interplanar spacing for the $j^{\rm th}$ family of planes in the equilibrium unit cell structure of component $i$. Each peak is represented by a Gaussian function, with $\alpha_j$ being the width of the peak, $j$. The $\{\alpha_j\}$ have been shown in Ref. \cite{Greenwood11} to set the elastic and surface energies, as well as
their anisotropic properties.

It is noted that the $\mbfk = 0$ mode of all correlation functions is essentially zero. In principle, as discussed above, the $\mbfk=0$ mode of these correlation functions can have their effects implicitly reflected through local coefficients in the free energy. In the case of a pure material, a nonzero peak height at $\mbfk = 0$ in the correlation function merely shifts the local free energy at densities away from the reference density, however the stability of equilibrium structures is typically unchanged~\cite{Greenwood11}. The situation is similar for alloys, where the $\mbfk = 0$ mode will have a negligible contribution for phases that remain relatively close to the reference density.  Deviations of phases away from the reference density will be manifested in the average density dimension of the phase diagram. Here, it is assumed that the average density $n_o=0$ to simplify the demonstration of the model. Of course, the more complex situations where both the concentration and average density need to be modelled can be treated by adding suitable $\mbfk=0$ contributions, or by choosing the appropriate coefficients in the bulk free energy. Thus, without loss of generality, we will assume no additional constant to the correlation function $\hat{C}_{2j}^{ij}$ here.

\subsection{Ternary Dynamics}
For the case of 3-component alloys, the dynamical equations of motions in Eqs.(\ref{dynamics-n})-(\ref{dynamics-c}) reduce to
\beq
\frac{\partial n}{\partial t} \!=\! M_n \nabla^2 \Biggl\{ \! n-\eta \frac{n^2}{2} + \chi \frac{n^3}{3} + \omega\Delta F_{\text {mix}}^{\text{ter}}-C_{eff}^{\text{ter}}\,n\!\! \Biggr\},
\label{3-dynamics-n}
\eeq
\beq
\frac{\partial c_A}{\partial t} \!=\! M_{A} \nabla^2  \Biggl\{ \! \omega(n+1)\frac{\delta \Delta F_{\text {mix}}^{\text{ter}}}{\delta c_A}
-\frac{1}{2}n\frac{\delta C_{eff}^{\text{ter}}}{\delta c_A}\,n - \kappa_A\nabla^2c_A \!\! \Biggr\}, \nonumber
\eeq
\beq
\frac{\partial c_B}{\partial t} \!=\! M_{B} \nabla^2  \Biggl\{\! \omega(n+1)\frac{\delta \Delta F_{\text {mix}}^{\text{ter}}}{\delta c_B}
-\frac{1}{2}n\frac{\delta C_{eff}^{\text{ter}}}{\delta c_B}\,n - \kappa_B\nabla^2c_B \!\! \Biggr\}, \nonumber
\eeq
where $M_n$, $M_{c_{A}}$ and $M_{c_{B}}$ are dimensionless mobility coefficients for density and compositions fields. They are set to 1 here, since it is the intent of this paper to introduce the model and its physical features.

\subsection{Equilibrium Properties}
\label{ternary-eqm-properties}
Ternary equilibrium is defined by co-existence of bulk phases, e.g. solid$_{\alpha}$-solid$_{\beta}$, liquid-solid$_{\alpha}$-solid$_{\beta}$, etc. The governing properties, e.g. partitioning, of such an equilibrium state can be determined from standard thermodynamic minimization methods. In general, for $3$-component alloys, free energy minimization is defined by a common plane tangent to the free energy wells of any two or three coexisting phases.  This construction is a geometrical representation of the statement that the chemical potentials and grand potentials of any two phases are equal with respect
to each component. Here, we construct isothermal ternary phase diagrams by examining all combinations of phase coexistence (e.g., solid$_\alpha$-liquid, solid$_\alpha$-solid$_\beta$, etc.). Procedures for calculating phase diagrams for PFC models are well-documented  \cite{Elder02,Elder07,Greenwood10,GreenwoodOfori11} and the approach used here will only be summarized.

For solid phases, the density field,  which varies on atomic length scales, is approximated using a multi-mode approximation given by
\beq
n_i(\mbfr~)=\sum_{j=1}^{N_i}A_j\sum_{l=1}^{N_j}\exp\left({\frac{2\pi}{a_i} i \mbfk_{l,j} \cdot \mbfr}\right),
\label{Density}
\eeq
where $a_i$ is the lattice spacing of the solid phase $i$ and $N_i$ denotes the number of mode families (families of planes) in the unit cell of phase $i$,
$A_j$ is the amplitude associated with the $j^{\rm th}$ family of planes. Each mode contains $N_j$ reciprocal lattice peaks, enumerated by the index $l$. Strictly speaking, there is a distinct amplitude, $A_{l,j}$, for each reciprocal lattice peak. However, for the purposes of simplifying the construction of phase diagrams (i.e., working with the fewest number of variables to minimize), they are assumed constant leading to $A_j$. Each index $l$ in the family $j$ has a corresponding reciprocal lattice vector $\mbfk_{l,j}$, normalized by the lattice spacing.

Substituting Eq.~(\ref{Density}) into Eq.~(\ref{3-simplified-Energy}), and integrating over one unit cell, the free energy can be calculated for each phase as a function of $c_A$, $c_B$ and the amplitudes $A_j$. Since amplitudes are non-conserved fields, the resulting free energy is then minimized with respect to each $A_j$ \cite{GreenwoodOfori11}. The result is substituted back into the free energy. After this procedure, we are left with a free energy landscape
${\cal F}_{sol}^{\text{ter}}(c_A,c_B)$, where ${\cal F}_{sol}^{\text{ter}}$ represents an amplitude-minimized solid free energy. In keeping with the discussion of the previous sections, we assume that the average density of all phases is close to the reference density, i.e., $n_{o}=0$. For the liquid phase, the free energy ${\cal F}_{liq}^{\text{ter}}(c_A,c_B)$ is trivially computed by setting all $A_j=0$
\footnote{For a more sophisticated treatment where phase density is allowed to vary, Eq.~(\ref{Density}) can be replaced by
$n_i(\mbfr~)=n_o^i +\sum_{j=1}^{N_i}A_j\sum_{l=1}^{N_j}\exp\left({\frac{2\pi}{a_i} i \mbfk_{l,j} \cdot \mbfr}\right)$.
The amplitude-minimized free energies then become of the form ${\cal F}_{i}^{\text{ter}}(c_A,c_B,n_o^i)$ for phase $i$. Equilibrium is then found via a ``common hyperplane construction'' that solves simultaneously for the concentrations and densities, a formidable problem. See appendix of Ref.~\cite{Provatas10} for this formalism derived for binary alloys.}.

With the free energy landscapes of liquid and solids, the phase boundary lines between a combination of phases at a given temperature parameter, $\sigma$, are computed by solving the following set of equations simultaneously,
\begin{align}
&\mu_{c_A}^{I}=\mu_{c_A}^{J}\nline
&\mu_{c_B}^{I}=\mu_{c_B}^{J}\nline
&\Omega^{I}=\Omega^{J},
\label{coex}
\end{align}
where the last of these implies,
\begin{align}
f^{I}-\mu_{c_A}^{I}c_{A}^{I}-\mu_{c_B}^{I}c_{B}^{I}=f^{J}-\mu_{c_A}^{J}c_{A}^{J}-\mu_{c_B}^{J}c_{B}^{J}.
\end{align}
The superscripts $I$ and $J$ denote any two phases in equilibrium (e.g. liquid-solid$_\alpha$), respectively. The expressions $\mu_{c_A}^I=\partial f^I/\partial {c_A}$ and $\mu_{c_B}^I=\partial f^I/\partial {c_B}$ are the  chemical potentials of phase $I$ with respect to the concentrations $c_A$ and $c_B$, respectively, with analogous definitions for $\mu_{c_A}^J$ and $\mu_{c_B}^J$. The expressions $\Omega^{I}$ and $\Omega^{J}$ are the grand potentials of phases $I$ and $J$, respectively. See Appendix~\ref{phase-diag-calc} for further details on calculating phase diagrams. The set of conditions in Eq.~(\ref{coex}), along with Eq.~(\ref{lever-coex}) defining the average concentration, can be solved to find the four equilibrium concentrations (two per phase) defining coexistence on a given tie line.

\subsubsection{Generic Ternary Eutectic Alloy}
\label{ternary-A-B-C-diagram}
\begin{figure*}[htbp]
\resizebox{6.in}{!}{\includegraphics{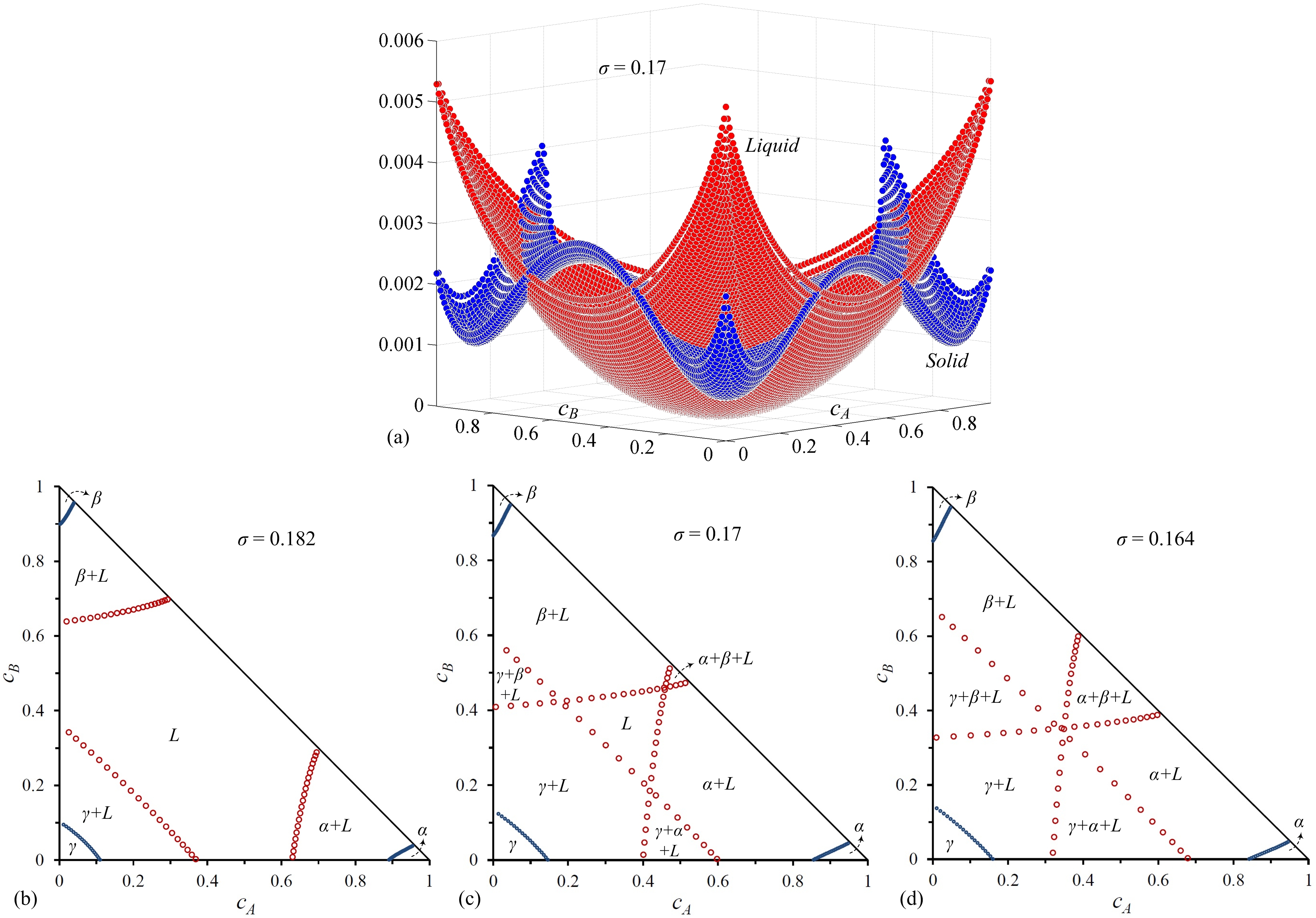}}
\caption{(Colour online) Ternary eutectic system: (a) Solid and liquid energy landscapes of a square-square-square ($A$-$B$-$C$) system at temperature parameter $\sigma=0.17$. Corresponding phase diagrams at temperatures (b) $\sigma=0.182$, (c) $\sigma=0.17$ and (d) $\sigma=0.164$. The parameters for ideal free energy and entropy of mixing were $\eta=1.4$, $\chi=1$, $\omega=0.005$, while reference concentrations were $c_A^o=0.333$ and $c_B^o=0.333$. Widths of the correlations peaks are taken $\alpha_{11}=0.8$ and $\alpha_{10}=\sqrt{2}\alpha_{11}$ for all phases (required for isotropic elastic constants in a solid phase with square symmetry~\cite{GreenwoodOfori11}). The peak positions for the given structures are $k_{11A}=(81/38)\pi$ and $k_{10A}=\sqrt{2}k_{11A}$ for $\alpha$, $k_{11B}=(54/29)\pi$ and $k_{10B}=\sqrt{2}k_{11B}$ for $\beta$ and $k_{11C}=2\pi$ and $k_{10C}=\sqrt{2}k_{11C}$ for $\gamma$. The effective transition temperatures are set to $\sigma_{Mj}=0.55$ for all family of planes in all phases. The concentrations on the isothermal phase diagrams are read in a Cartesian coordinate system. }
\label{fig:SL-PhaseDiagram}
\end{figure*}

A first example of the equilibrium properties of the ternary XPFC model are demonstrated for a system where all three components ($A,B,$ and $C$) are structurally similar, differing only in their equilibrium lattice spacings. Here two-dimensional (2D) square symmetry is assumed as the equilibrium structure for each pure component, which in this context implies that all $\hat{C}_2^{ii}$ have the same number of peaks, with the corresponding ratios of their positions in reciprocal space being the same. However, each structure is differentiated by the absolute positions ($k_j$) of each peak. Though it has not been done in this initial work, by adjusting the widths ($\alpha_j$) of each peak, each element can also be differentiated by different elastic and surface energies. The full list of parameters used to construct the phase diagrams in this subsection are listed in the caption of Fig.~\ref{fig:SL-PhaseDiagram}.

Allowing all three components to have square structural symmetry, at sufficiently low temperature we can construct a bulk solid free energy landscape describing multiple solid phases, described by an effective lattice parameter ($a^{\text{ter}}$) that is a weighted average of the individual lattice parameters of all three components, using the interpolation functions of Eq.(\ref{interp-func}), namely, $a^{\text{ter}}=a_AX_A+a_BX_B+a_CX_C$. This leads to the solid-liquid free energy landscape in Fig.~\ref{fig:SL-PhaseDiagram}(a) for $\sigma=0.17$, where the values of all other parameters are specified in the figure caption. The corresponding isothermal phase diagram is illustrated in Fig.~\ref{fig:SL-PhaseDiagram}(c), which is constructed form the coexistence lines calculated between the liquid phase and the different solid-solution phases, using the set of conditions in Eq.~(\ref{coex}). Figure~\ref{fig:SL-PhaseDiagram}(b) shows an isothermal cut at a higher temperature, i.e., $\sigma=0.182$, depicting an increased region where the bulk liquid is stable compared to the solid phases. At sufficiently low temperature, the free energy admits eutectic coexistence of three phases. We construct an isothermal cut right above the eutectic temperature, i.e., at $\sigma=0.164$, shown in Fig.~\ref{fig:SL-PhaseDiagram}(d). The corresponding concentrations $c_A$ and $c_B$ in Fig.~\ref{fig:SL-PhaseDiagram}, are given as fractions, where unity represents pure $A$ or $B$, respectively, along each axis of the phase diagram.

\subsubsection{Simplified Al-Cu-Mg Type Alloy}
\label{ternary-Al-Cu-Mg-diagram}
\begin{figure*}[htbp]
\resizebox{6.in}{!}{\includegraphics{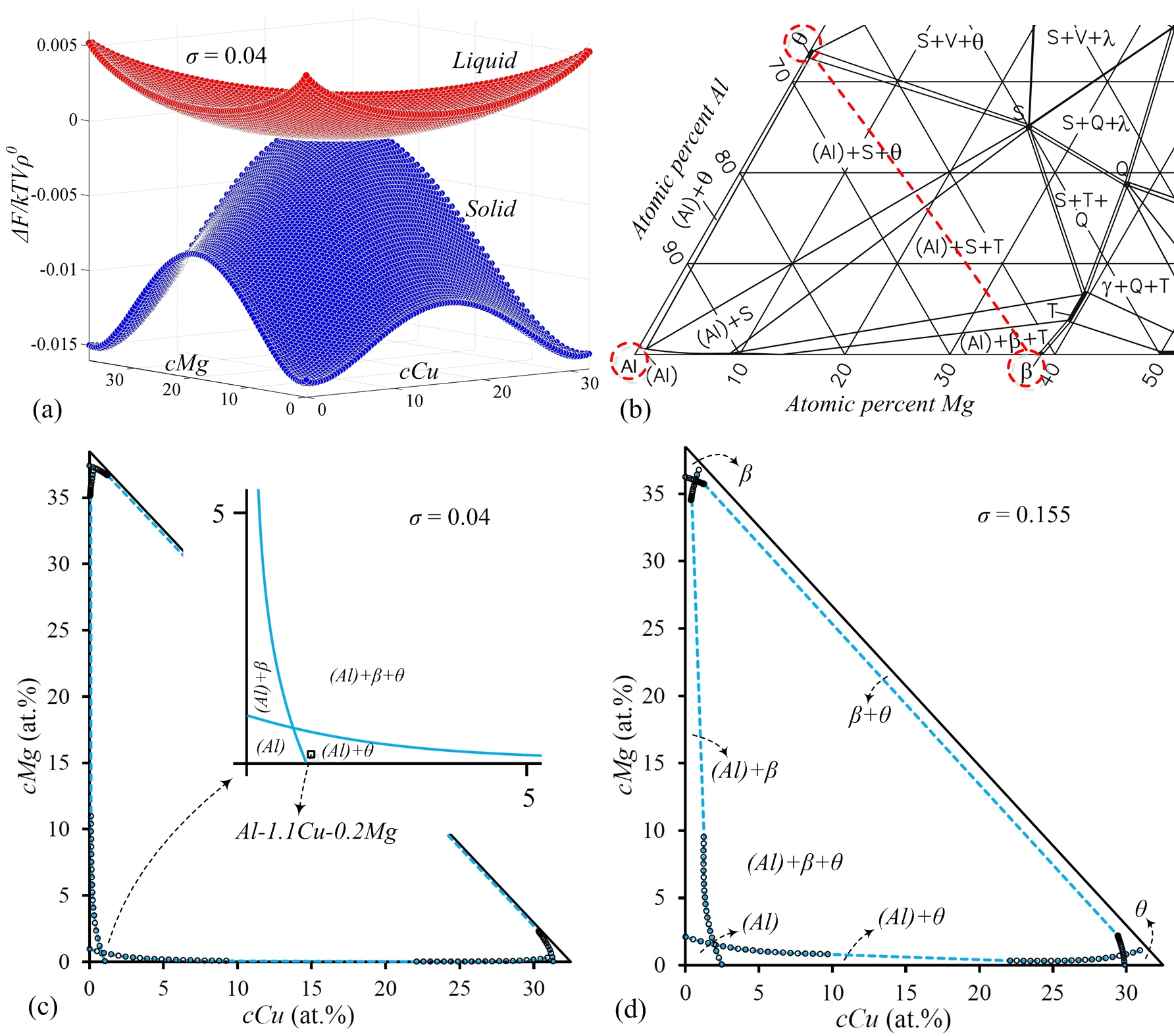}}
\caption{(Colour online) Al-Cu-Mg Phase diagram:  (a) Solid and liquid energy landscapes of a square-square-square ((Al)$-\beta-\theta$) system at temperature $\sigma=0.04$, (b) The Al-rich side of an isothermal cut (at $400^\circ C$) from the experimental phase diagram of the Al-Cu-Mg system taken from Ref.~\cite{Raghavan07}. Dashed circles mark the regions of the Al-rich (Al), Cu-rich ($\theta$) and Mg-rich ($\beta$) regions considered for reconstruction by the model phase diagram. Reconstructed phase diagrams at temperatures (c) $\sigma=0.04$ and (d) $\sigma=0.155$. The parameters for ideal free energy and entropy of mixing were $\eta=1.4$, $\chi=1$, $\omega=0.005$, $c_{Cu}^{o}=0.333$ and $c_{Mg}^{o}=0.333$. Widths of the correlations peaks are $\alpha_{11}=0.8$ and $\alpha_{10}=\sqrt{2}\alpha_{11}$ for all phases. The peak positions are $k_{11(Al)}=2\pi$, $k_{10(Al)}=\sqrt{2}k_{11(Al)}$, $k_{11\theta}=(2.0822)\pi$, $k_{10\theta}=\sqrt{2}k_{11\theta}$, $k_{11\beta}=(1.8765)\pi$ and $k_{10\beta}=\sqrt{2}k_{11\beta}$. For all family of planes, $\sigma_{Mj}=0.55$, in all phases. The maxima in concentrations $c_{Cu}$ and $c_{Mg}$ are rescaled from unity, 1, to correspond to the Cu and Mg-content in the $\theta$-phase and $\beta$-phase given by the experimental phase diagram, i.e., $\approx 32.5$ and $\approx 38.5$ $at.\%$, respectively. The concentrations on the isothermal phase diagrams are read in a Cartesian coordinate system. }
\label{fig:SS-PhaseDiagram}
\end{figure*}

The parameters of the ternary XPFC model can be chosen to produce sections of experimental phase diagrams qualitatively, as in the work of Fallah {\it et al.}~
\cite{Fallah12b}, where the present ternary model is used to model precipitation in a 2D representation of the Al-Cu-Mg system. Here, we demonstrate how the equilibrium properties of a portion of the Al-rich (simplified) part of the Al-Cu-Mg phase diagram can be described quantitatively by the ternary XPFC model. An experimental phase diagram at $400^\circ C$ is shown in Fig.~\ref{fig:SS-PhaseDiagram}(b), taken from Ref.~\cite{Raghavan07}.

Consider the part of the phase diagram for (Al)$-\beta$-$\theta$ outlined by the red dashed line and circled solid phases in the experimental phase diagram shown in Fig.~\ref{fig:SS-PhaseDiagram}(b), and ignoring the (Al)+S and (Al)+$\beta$+T phase regions. In the dilute-Mg region, a eutectic transition occurs between the Al-rich, (Al)-fcc phase, and an intermediate phase $\theta$ which has a tetragonal crystal structure. The eutectic system of (Al)-$\theta$ has a small solubility for Mg, however past the maximum solubility limit, there exists other intermediate phases terminating at the cubic $\beta$-phase. The equilibrium lattice constants (and thus the positions of the reciprocal space peaks) of $\theta$ and $\beta$ phases are determined by interpolating between those of Al with $32.5$\,at.$\%$Cu, and Al with $38.5$\,at.$\%$Mg, respectively. For simplicity, we assume a square structural symmetry for all three equilibrium phases, and like the preceding section, the effective lattice constant is interpolated by weighting by local solute compositions, $c_{Cu}$ and $c_{Mg}$. The parameters ($\eta, \chi, \omega$), along with the peak widths $\alpha_j$ are chosen to give a satisfactory mapping of the solubility limits of the (Al)-phase for Cu and Mg to those in the experimental phase diagram of Fig.~\ref{fig:SS-PhaseDiagram}(b), for a range of temperature parameters ($\sigma$). The full list of parameters used to construct the phase diagrams in this subsection are listed in the caption of Fig.~\ref{fig:SS-PhaseDiagram}.

Figure~\ref{fig:SS-PhaseDiagram}(a), shows the free energy landscape for the solid at $\sigma=0.04$. Figure~\ref{fig:SS-PhaseDiagram}(c) shows the corresponding isothermal phase diagram  at $\sigma=0.04$, where the inset shows a zoomed in image of the Al-rich corner. Comparing the inset with the experimental phase diagram, reasonable agreement is evident between the calculated and the experimental phase diagram sections. Figure~\ref{fig:SS-PhaseDiagram}(d) shows the isothermal phase diagram for $\sigma=0.155$. At this higher temperature (still below the eutectic), there is an increase in the solubility limits of the phase boundaries. Section~\ref{ternary-clustering} will use  this phase diagram to demonstrate solid-state precipitation.

\section{Applications}
\label{ternary-apps}
The binary XPFC approach was previously demonstrated as a tool with which to model the role of defects and elasticity in structural phase transformations that operate over diffusive time scales. Further to these capabilities, the ability to have multi-component interactions between solute atoms and defects now makes it possible to examine much more complex interactions of the above atomic-scale effects with different solutes, and their diffusion. This capability opens a myriad of possibilities for applications for microstructure engineering in materials. This section showcases some applications of the XPFC multi-component model presented in this work. In particular, using the phase diagrams from the previous section, we demonstrate dendritic solidification and precipitation in the presence of ternary components. These phenomena are paradigms of microstructure evolution of relevance to materials engineering applications and are strongly influenced by diffusion of impurities, elastic strain, crystal anisotropy and defect structures.

\subsection{Dendritic Solidification}
\label{ternary-solidificaiton}
\begin{figure}[t]
    \centering
    \begin{tabular}{cccc}
    \includegraphics[width=1.in]{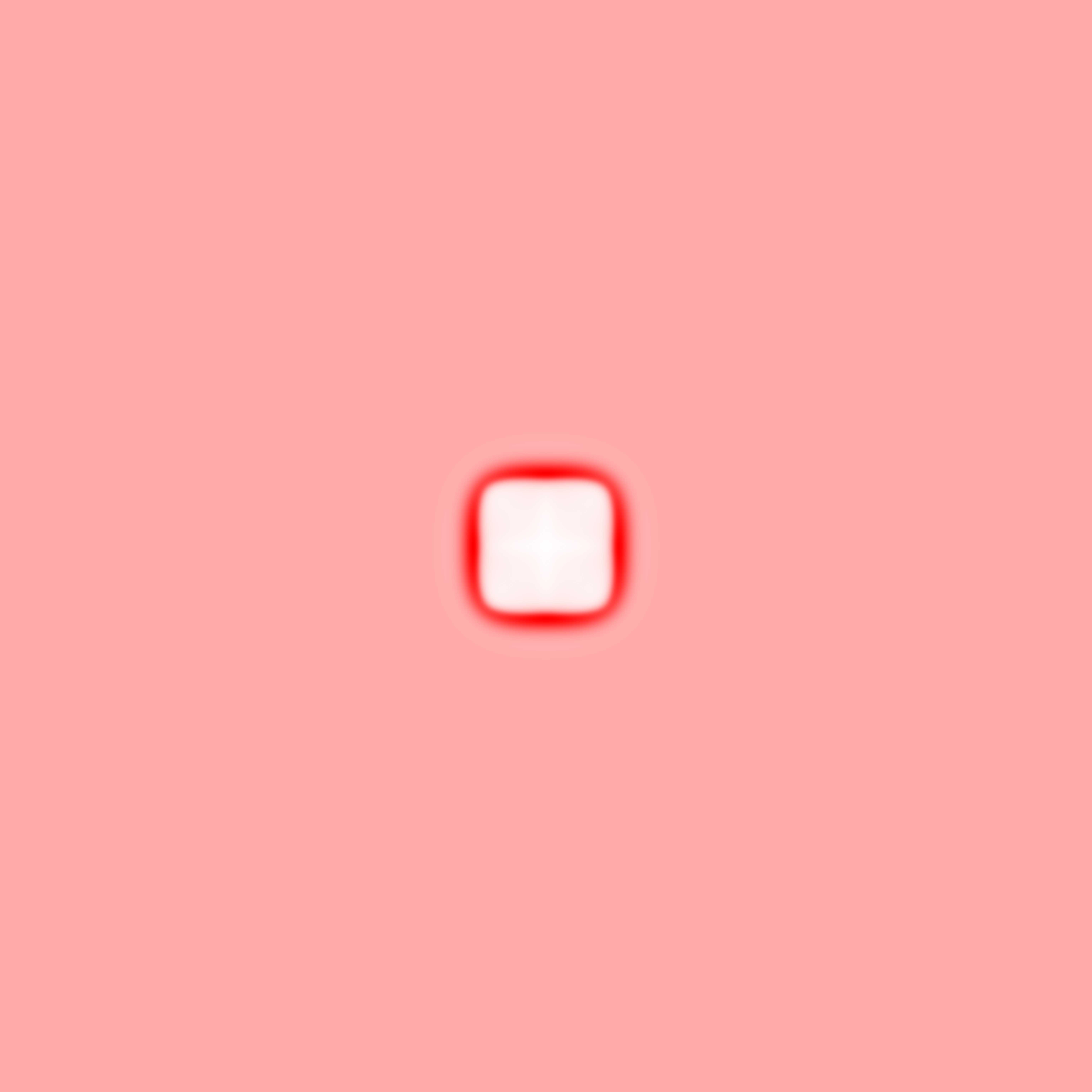}&\includegraphics[width=1.in]{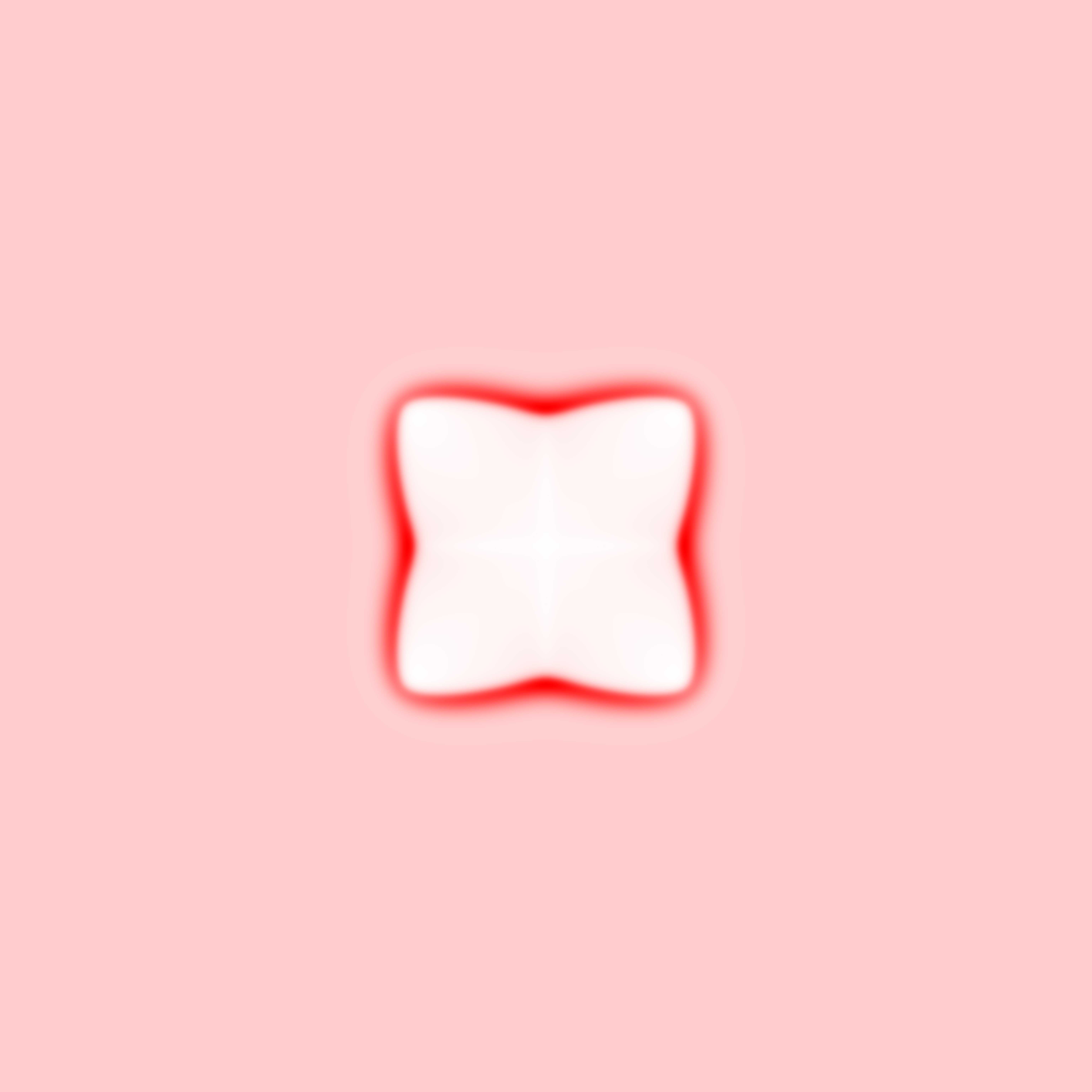}&\includegraphics[width=1.in]{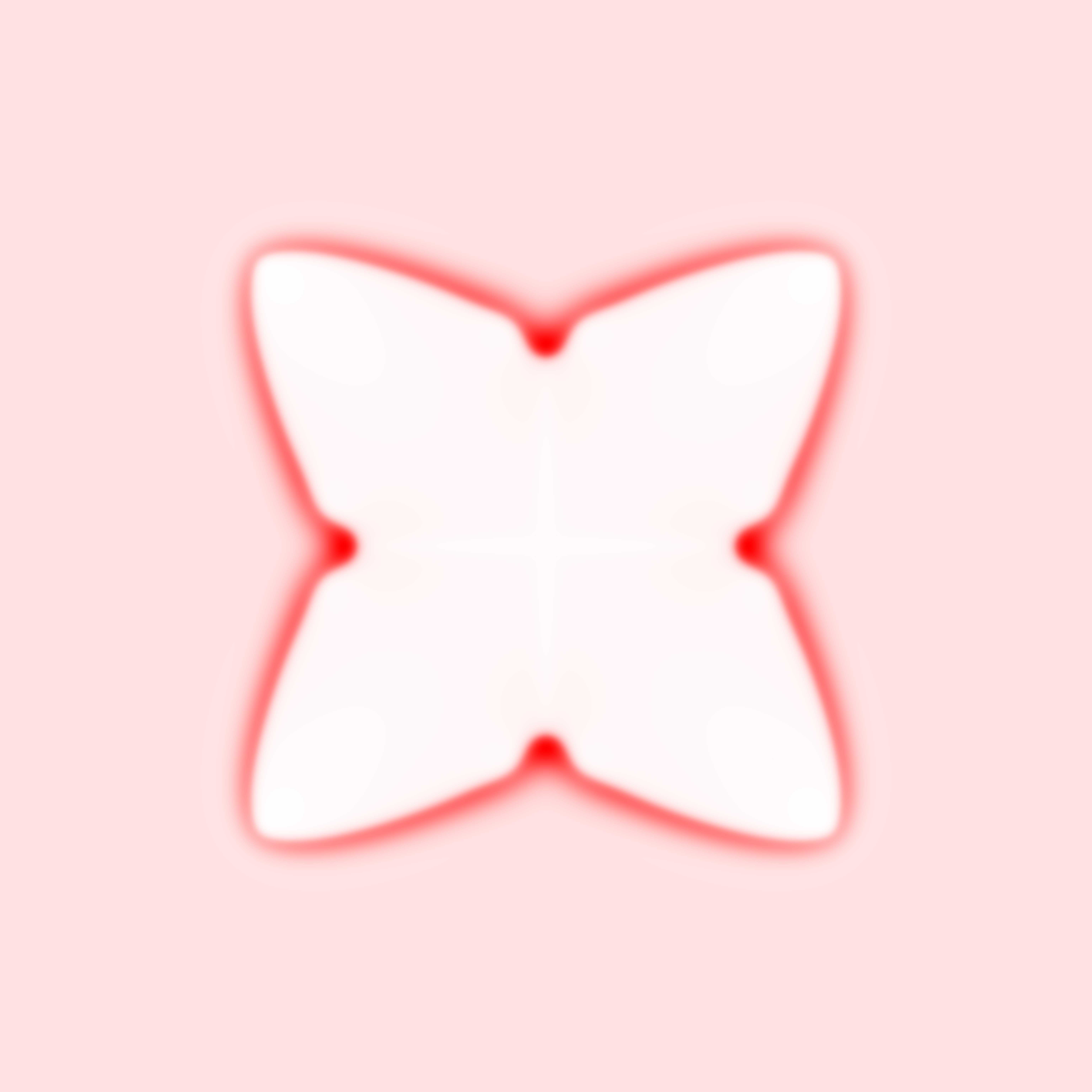}
    \\
    \includegraphics[width=1.in]{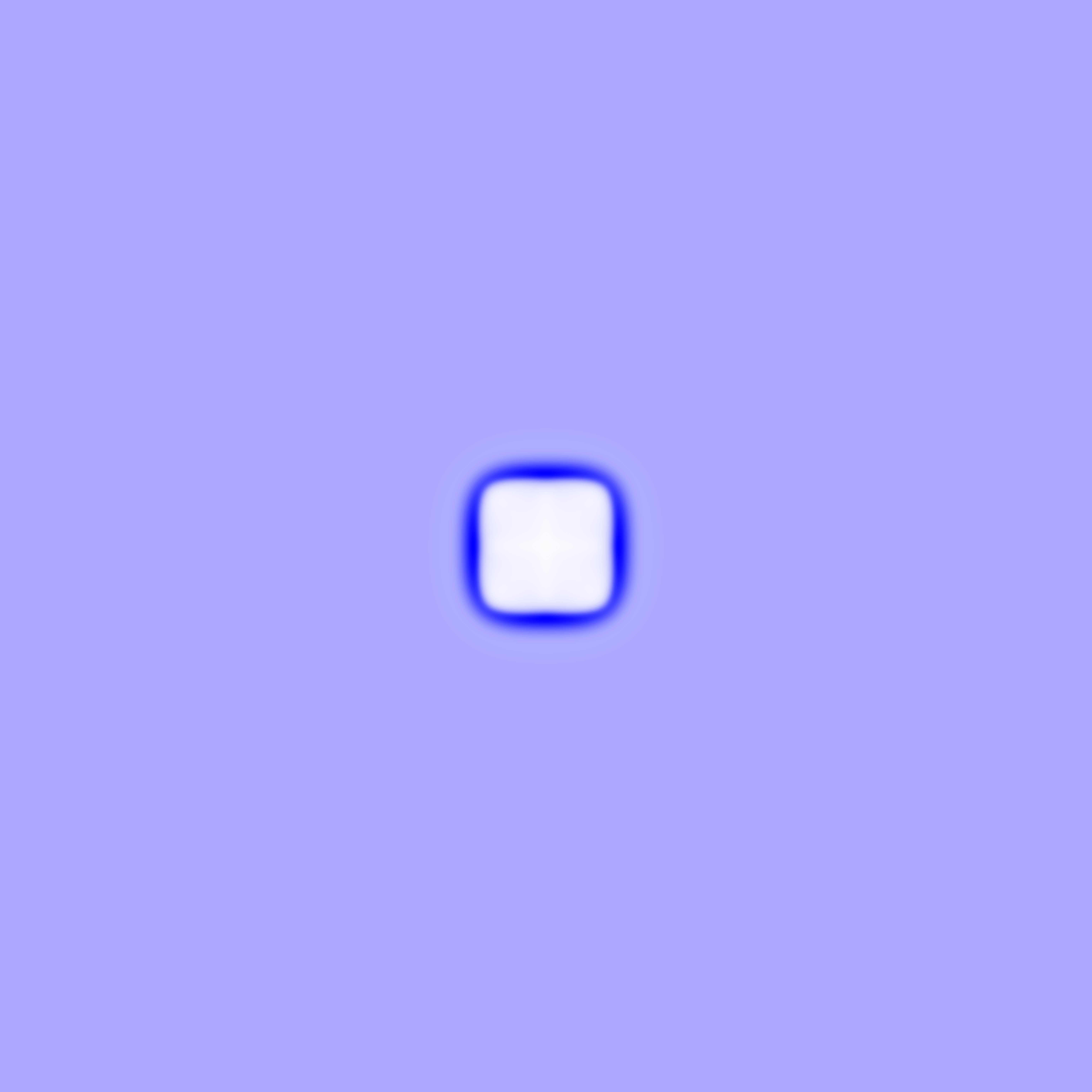}&\includegraphics[width=1.in]{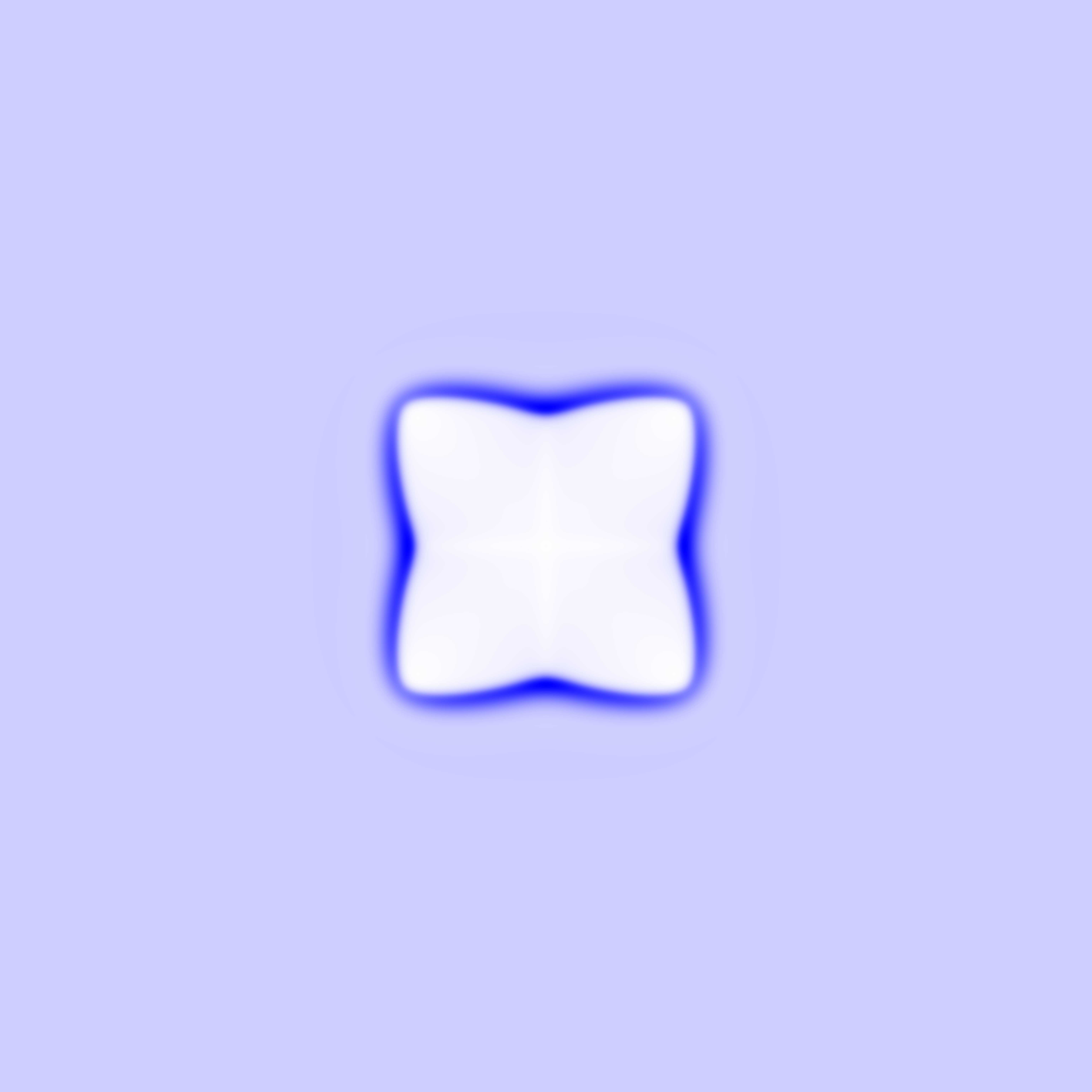}&\includegraphics[width=1.in]{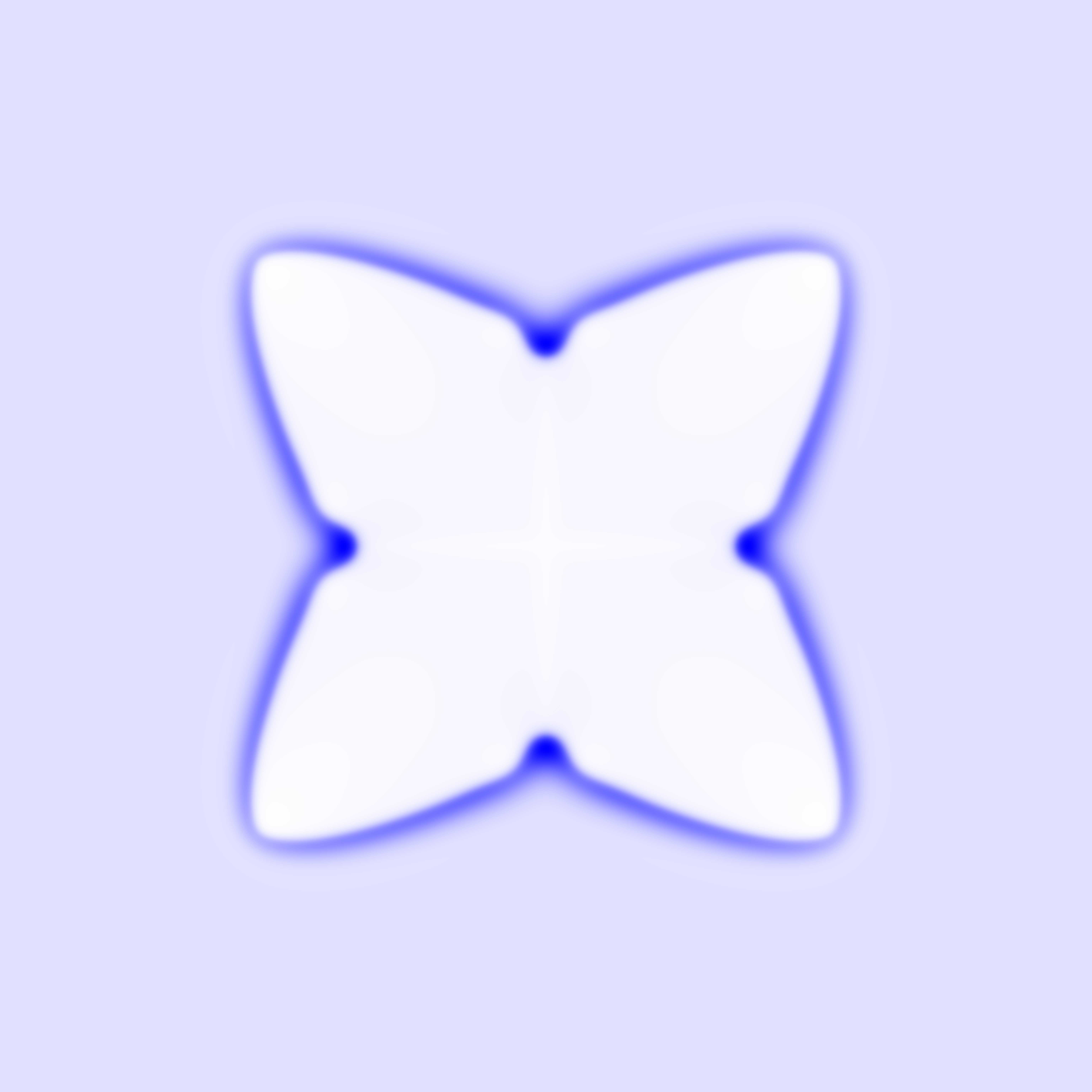}
    \\
    \includegraphics[width=1.in]{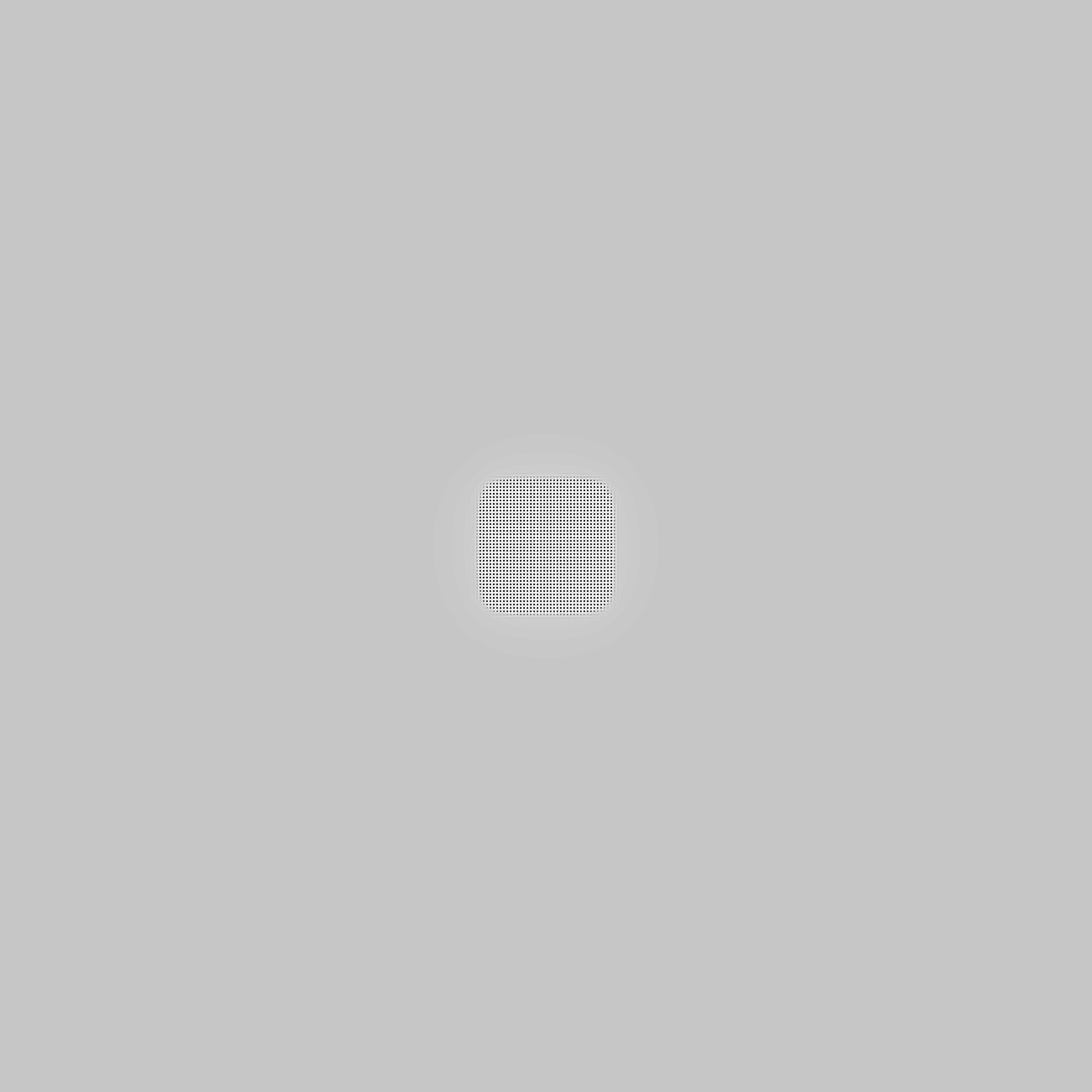}&\includegraphics[width=1.in]{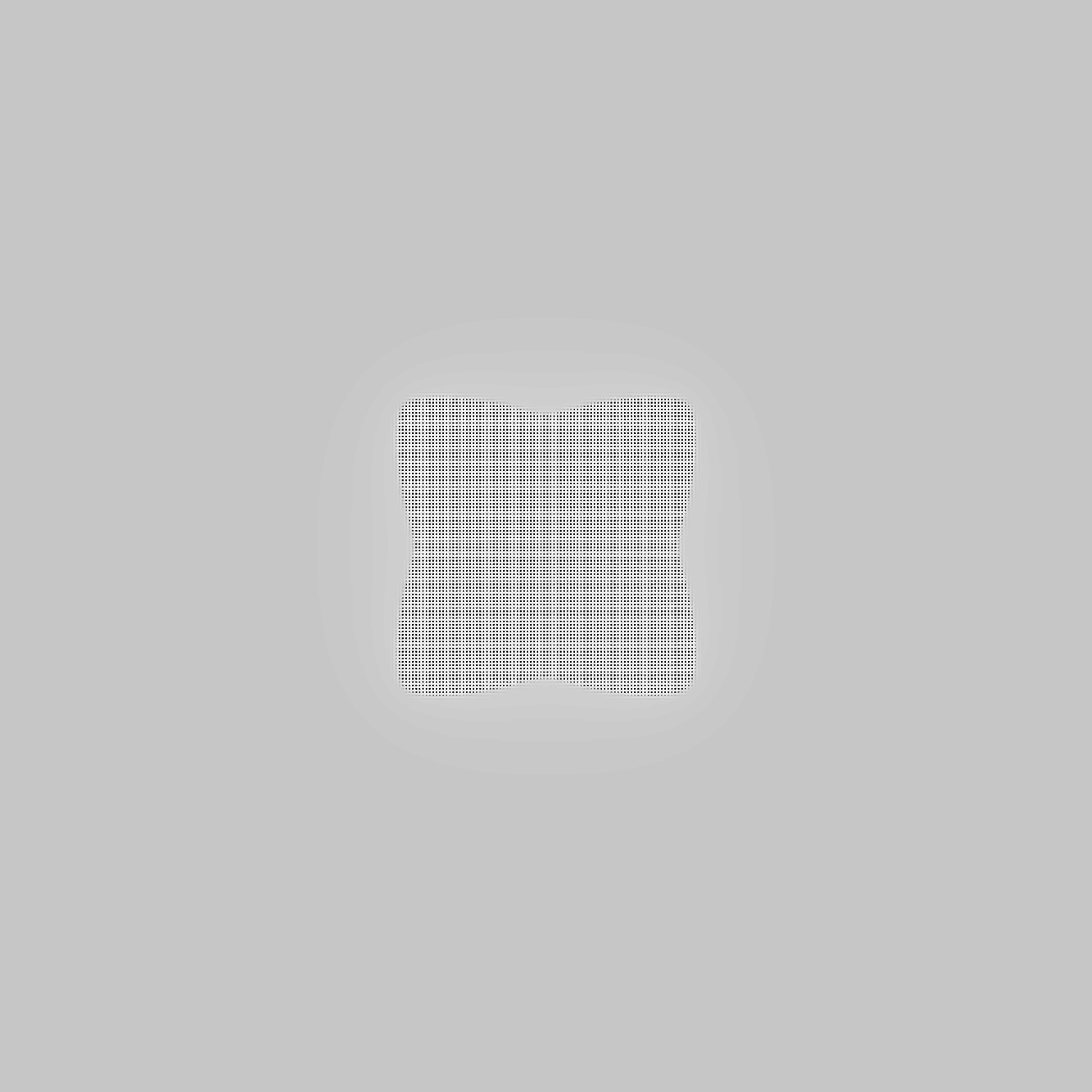}&\includegraphics[width=1.in]{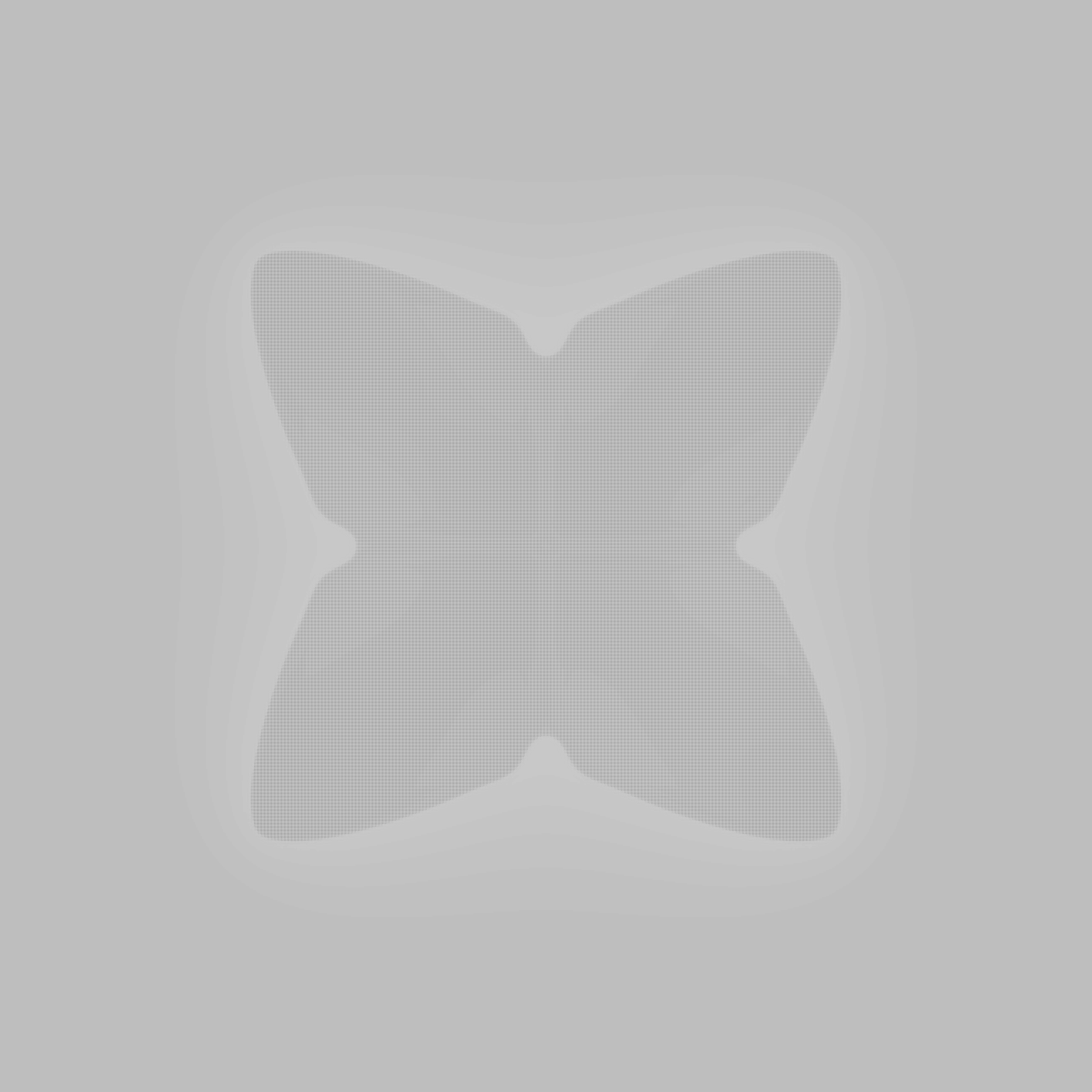}
    \\
    (a)&(b)&(c)
    \end{tabular}
    \caption[]{(Colour online) Early-time dendritic solidification in a ternary alloy, simulated using the phase diagram of Fig.~\ref{fig:SL-PhaseDiagram}(b). The quench temperature is $\sigma=0.182$ and the initial solute compositions are uniform and set to the alloy averages, $\bar{c}_{A}=0.1$ and $\bar{c}_{B}=0.1$. Each Column of images represents a different time during the simulation. The times shown are: (a) 1000 (b) 3000 and (c)7000 iterations. From bottom to top, each row displays the progression of $n$, $c_B$ and $c_A$, respectively, with $c_{A}$ plotted in the colour range from white (lowest concentration) to red (highest concentration) and $c_{B}$ is plotted in the colour range from white (lowest concentration) to blue (highest concentration).}
    \label{fig:dendrite}
\end{figure}
\begin{figure}[t]
    \centering
    \includegraphics[width=3.2in]{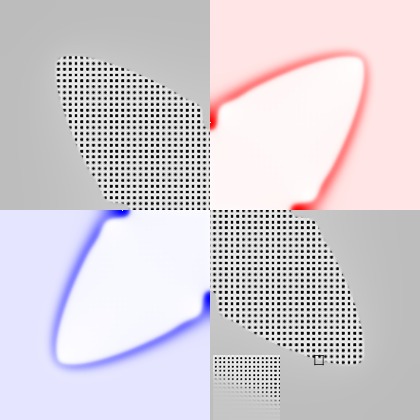}
    \caption[]{(Colour online) Dendritic solidification at time, $t=10000$, displaying the values of all three fields. The top left and bottom right quadrants show the density field, $n$. The inset on the bottom right, is a zoomed-in image of the rectangular area marked in black, revealing the structure of the underlying square lattice and density wave structure through the interface. Top right quadrant shows $c_{A}$, with colour range white (low concentration) to red (high concentration), while bottom left shows $c_{B}$, with colour range white (low concentration) to red (high concentration). Both show the high solute content at the interface. }
    \label{fig:4play}
\end{figure}
Dendritic solidification arises when a supercooled liquid is quenched into the solid-liquid coexistence part of the phase diagram. Figure~(\ref{fig:dendrite}) shows snapshots in time of a dendritic crystal in a ternary alloy. The simulation was done using the phase diagram in Fig.~\ref{fig:SL-PhaseDiagram}(b). Simulations were conducted in a 2D domain of size $768a\times768a$, where $a$ is the lattice spacing. A uniform grid spacing and discrete time of $\Delta x = 0.125$ (which makes for a domain of size $6144\times6144$ grid points) and $\Delta t =3$ were used and equations of motion, Eq.~(\ref{3-dynamics-n}), were solved semi-implicitly in Fourier space. The initial conditions consisted of a small circular seed of diameter, $d=8a$ of $\gamma$-phase, seeded in liquid at a temperature of $\sigma=0.182$. The initial concentration of solute components $A$ and $B$ was uniform in both phases and set to the values $\bar{c}_A=0.1$ and $\bar{c}_B=0.1$. Several time slices of the simulation domain, showing the fields ($n,c_{A}$ and $c_{B}$) at early tines, are shown in Fig.~\ref{fig:dendrite}.

As time progresses during the simulation, Fig.~\ref{fig:dendrite}(a)-(c), dendritic growth is evident. The crystal develops a characteristic 4-fold symmetry of the underlying square crystal structure, produced with the correlation function for the given pure component of the $\gamma$-phase. The top two rows show the time evolution of the concentration fields (from left to right), $c_{A}$ and $c_{B}$, respectively, indicating the interface boundary layer for each component. Both solutes, $A$ and $B$, reach their maximum solute content at the interface of the growing dendrite, in agreement with the solute rejection mechanism of crystal growth. The bottom row shows the evolution of the density. There is also evidence of the associated density jump at the interface between solid and liquid phases as depicted by the light halo like region around the interface. Figure~\ref{fig:4play} shows a composite view of  the dendrite at later time, highlighting in each quadrant one of the three fields. This simulation depicts multiple diffusing species, density changes and surface tension anisotropy. In a larger numerical domain (where multiple dendrites can be grown), grain boundaries would also naturally emerge. It is noteworthy that these physical ingredients  arise self-consistently and are very straightforward to simulate numerically. We also note that side-branching of the growing dendrite is not observed in Fig.~\ref{fig:dendrite} due to the size of the simulation domain and the exclusion of thermal noise in the dynamical equations.

\subsection{Solute Clustering and Precipitation}
\label{ternary-clustering}
Many properties of engineering alloys are typically attained through downstream processing following solidification. These downstream processes typically involve either thermo-mechanical manipulations or heat treatment of the as-cast microstructure. One of the most important aims is to induce certain phase transformations in the as-cast primary solid matrix to help strengthen alloys, a process known as precipitation hardening. In this subsection we demonstrate this process using the ternary XPFC model developed in this work. In particular, we illustrate the initial stages of a heat treatment process leading to solute clustering, the precursor stage of precipitation in Al-Cu-Mg alloys.  The details of this process have been reported elsewhere \cite{Fallah12b}.

\begin{figure*}[htbp]
\resizebox{6.in}{!}{\includegraphics{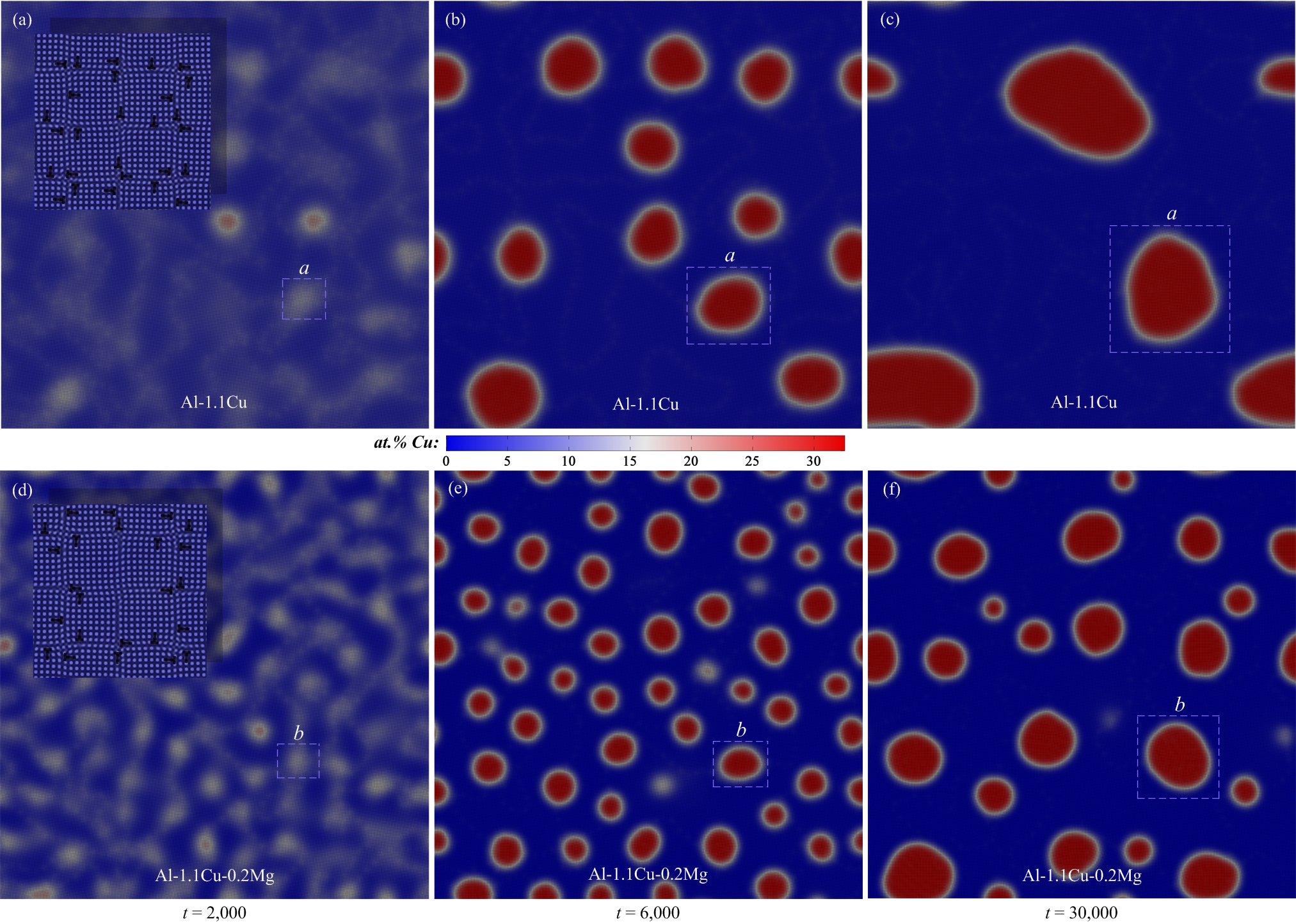}}
\caption{(Colour online) Time evolution of clusters in solutionized/quenched (a)-(c) Al-1.1Cu and (d)-(f) Al-1.1Cu-0.2Mg alloys at $\sigma=0.04$. The insets in (a) and (d) show the initial distorted/damaged single-phase structures, with dislocations clearly marked, for each set of simulations.}
\label{fig:Clustering}
\end{figure*}
Solute clustering/early-stage precipitation simulations were preformed using the equilibrium properties calculated for the (Al)$-\beta$-$\theta$ system in Figs.~\ref{fig:SS-PhaseDiagram}(c) and \ref{fig:SS-PhaseDiagram}(d). Simulations were performed on a 2D rectangular mesh with grid spacing $\Delta x=0.125$ and time step $\Delta t=10$. Dynamical equations were solved semi-implicitly in Fourier space. Initial conditions consisted of distorted single-phase structures, through the introduction of a uniform distribution of dislocations, and a uniform composition everywhere of $c_{Cu}=1.1$ and $c_{Mg}=0.2$ at.\%. All simulations were initially solutionized for some time at $\sigma=0.155$, following which they were quenched/aged at a temperature $\sigma=0.04$. During ageing, small clusters initially appear with higher Mg and/or Cu-content than that of the matrix. As time progresses, some of these clusters decrease in size and Cu-Mg-content, or vanish entirely. A few, however, stabilize, as shown by the typical stabilized clusters ``$a$'' and ``$b$'' in Fig.~\ref{fig:Clustering}(a)-(c) and (d)-(f) for Al-1.1Cu and Al-1.1Cu-0.2Mg alloys, respectively. In contrast, for either alloy, when we increase the ageing temperature within the single-phase (Al) region , e.g., $\sigma=0.145$, no clustering is observed and the initial distortions are removed from the matrix.

Experiments in quenched/aged Al-Cu and Al-Cu-Mg alloys~\cite{Ozawa70,Somoza02,Nagai01} have found increasing evidence that the interaction of ternary impurities and quenched-in defects such as dislocations \cite{Babu12} dynamically reduce the local nucleation barrier for precipitation at locations in the matrix. We have also found that the addition of Mg into an Al-1.1Cu alloy promotes clustering and refinement of the final microstructure, as seen in the simulation data of Fig.~(\ref{fig:Clustering}). The clustering phenomenon observed in these simulations can be attributed to the propensity for solute segregation to defects and surrounding areas to relieve stresses induced by the presence of said defects, in this case dislocations. As more solute aggregates to dislocations, the size of the cluster increases but the structural nature of the cluster also begins to approach that of the next nearest stable solid phase. As this process continues and the ever growing cluster attracts more solute, it creates additional stresses in the surrounding matrix. This in turn draws nearby dislocations to the cluster in attempts to relieve these additional stresses caused by solute accumulation. An extensive investigation of solute clustering mechanisms, in presence of quenched-in bulk crystal defects, has been done through a quantitative analysis of the system energetics in binary alloys in ~\cite{Fallah12a} and recently in ternary alloys, using the present model \cite{Fallah12b}.

\section{Summary}
\label{summary}

This paper reported a new phase field crystal model for structural phase transformations (XPFC) in multi-component alloys. The details of the model derivation were discussed. A simplified version of the model was specialized for ternary alloys  and its equilibrium properties were shown. The dynamics of the model were demonstrated on two phenomena of relevance to microstructure evolution in materials science.

This is the first multi-component PFC model, and as such is able to capture the complex kinetics of solidification and elastic and plastic effects on solid state processes, such as clustering and precipitate growth. This model has been used in a separate work \cite{Fallah12b}  to support recent experiments on the elusive mechanisms of the early stages of clustering and precipitation.

The phase field crystal methodology was introduced to create a bridge between the atomic and traditional phase field regimes. As a relatively novel method, many works in this area of materials science are working to validate the physics of PFC models. As the first phase field crystal model for $N$-component alloys, this work has demonstrated some important thermodynamic and kinetic  properties of the model. Moreover, aside from the model's quantitative and self-consistent nature, it is particularly simple to operate numerically. It is expected that this model can thus be used to elucidate the role of multiple solutes in phenomena governed by atomic-scale elasticity and defects operating on diffusional time scales.

\acknowledgements{We thank the Natural Science and Engineering
Research Council of Canada (NSERC) and Ontario Ministry of Research and In-
novation (Early Researcher Award Program) for financial support, and the CLUMEQ Supercomputing facility of Compute Canada.}

\appendix
\section{Long Wavelength Limit}
\label{long-wavelength-limit}
In section~\ref{simplified-PFC-multi-functional}, we reduced the free energy in Eq.~(\ref{multi-full-energy}) into a simplified multi-component XPFC energy functional. In the process of doing this, we simplified terms by considering the long wavelength limit where the concentration varies much more slowly than the density field. This  Appendix details the steps of how some terms of Eq.~(\ref{multi-full-energy}) can be simplified to derive the simplified free energy functional in Eq.~(\ref{simplified-Energy}).

\subsection{Terms Coupling Product of $c_i$ and $c_j$ with $C_2^{ij}$}
We begin first with terms involving a coupling of two concentration fields with a correlation function. As a concrete example, consider the term
\begin{align}
\label{grad-terms-1}
{\cal G} =-\frac{1}{2}\sum_{i,j}^N \int d\mbfr\,c_{i}(\mbfr)\,\int d\mbfr^{\prime}\, C_2^{ij}(|\mbfr-\mbfr^{\prime}|)c_{j}(\mbfr^{\prime}),
\end{align}
in Eq.~(\ref{multi-full-energy}), where we have used the more explicit notation for clarity. (The other terms follow analogously.) To proceed, we rewrite the correlation function in a Fourier series of the form,
\begin{align}
\label{C2-fourier}
C_2^{ij}(|\mbfr-\mbfr^{\prime}|) = \int d\mbfk\,\hat{C}_2^{ij}(|\mbfk|)e^{\mbfi\mbfk\cdot\mbfr}e^{-\mbfi\mbfk\cdot\mbfr^{\prime}}.
\end{align}
Substituting Eq.~(\ref{C2-fourier}) into Eq.~(\ref{grad-terms-1}) yields,
\begin{align}
\label{grad-terms-2}
\tilde{{\cal G}} =-\frac{1}{2}\sum_{i,j}^N \int d\mbfr\,c_{i}(\mbfr)\int d\mbfk\,\hat{C}_2^{ij}(|\mbfk|)\hat{c}_{j}(\mbfk)e^{\mbfi\mbfk\cdot\mbfr},
\end{align}
where we define
\begin{align}
\hat{c}_{j}(\mbfk) \equiv \int d\mbfr^{\prime}\,c_{j}(\mbfr^{\prime})e^{-\mbfi\mbfk\cdot\mbfr^{\prime}}.
\end{align}
Considering the long wavelength limit, we take a Taylor series expansion of the correlation function in powers of $\mbfk^2$ around $\mbfk=0$. This results in,
\begin{align}
\tilde{{\cal G}} &=\label{grad-terms-3} \\
&-\!\frac{1}{2}\sum_{i,j}^N \int \! d\mbfr\,c_{i}(\mbfr) \!\! \int \!\! d\mbfk\,\sum_{l=0}^{\infty}\frac{(-1)^{l}}{l!}(\mbfk^2)^{l}\frac{\partial^{l}\hat{C}_2^{ij}}{\partial (\mbfk^2)^{l}}\bigg|_{\mbfk=0} \!\!\!\!\!\!\! \hat{c}_{j}(\mbfk)e^{\mbfi\mbfk\cdot\mbfr}.
\nonumber
\end{align}
We note that to invoke the long wavelength limit, we could have also Taylor expanded the concentration, $c_{j}(\mbfr^{\prime})$, at $\mbfr^{\prime}=\mbfr$ as is done in Refs.~\cite{Wu07,Majaniemi09,Provatas10} or employed the multi-scale expansion used in Refs.~\cite{Elder10,Huang10}. All these methods, though different and require different mathematical treatments, are found to be equivalent. Retaining, to lowest order, terms up to order $l=1$, we have
\begin{align}
\label{grad-terms-4}
\tilde{{\cal G}} &=-\frac{1}{2}\sum_{i,j}^N \int d\mbfr\,c_{i}(\mbfr)\int d\mbfk\,\hat{C}_2^{ij}(|\mbfk|)\bigg|_{\mbfk=0}\hat{c}_{j}(\mbfk)e^{\mbfi\mbfk\cdot\mbfr}\nonumber\\
&+\frac{1}{2}\sum_{i,j}^N \int d\mbfr\,c_{i}(\mbfr)\int d\mbfk\,\mbfk^2\frac{\partial\hat{C}_2^{ij}}{\partial (\mbfk^2)}\bigg|_{\mbfk=0}\hat{c}_{j}(\mbfk)e^{\mbfi\mbfk\cdot\mbfr}.
\end{align}

Using the definition of the inverse Fourier transform, we recast Eq.~(\ref{grad-terms-4}) as
\begin{align}
\label{grad-terms-5}
\tilde{{\cal G}} &=-\frac{1}{2}\sum_{ij}^N\,\gamma_{ij}\, \int d\mbfr\,c_{i}(\mbfr)\,c_{j}(\mbfr)\nonumber\\
&+\frac{1}{2}\sum_{i,j}^N\, \kappa_{ij}\,\int d\mbfr\,c_{i}(\mbfr)\left(-\nabla^2\right){c}_{j}(\mbfr),
\end{align}
where we have used the following definitions,
\begin{align}
\label{gamma}
\gamma_{i j}\equiv \hat{C}_2^{ij}(|\mbfk|)\bigg|_{\mbfk=0}
\end{align}
and
\begin{align}
\label{kappa}
\kappa_{i j}\equiv \frac{\partial\hat{C}_2^{ij}}{\partial (\mbfk^2)}\bigg|_{\mbfk=0}.
\end{align}
It is thus clear that the first term in Eq.~(\ref{grad-terms-5}) will contribute terms that renormalize the coefficient of the $c_i^2$ terms in the entropy of mixing, if Eq.~(\ref{simplified-Energy}) were expanded about $c_i=c_i^o$.  In this work, the $\gamma_{ij}$ terms are neglected, and their role is subsumed in an
effective manner, for convenience, into the prefactor $\omega$ in Eq.~(\ref{simplified-Energy}). The second term in Eq.~(\ref{grad-terms-5}) can be recast into gradient energy terms analogous to those used in Cahn-Hilliard or Ginzburg-Landau theories. To do so, we perform integration by parts, yielding,
\begin{align}
\label{grad-terms-6}
\tilde{{\cal G}} &=-\frac{1}{2}\sum_{i,j}^N\,\gamma_{ij}\, \int d\mbfr\,c_{i}(\mbfr)\,c_{j}(\mbfr)\nonumber\\
&+\frac{1}{2}\sum_{i,j}^N \kappa_{ij}\int d\mbfr\,\nabla c_{i}(\mbfr)\cdot\nabla{c}_{j}(\mbfr).
\end{align}
Lastly, for clarity of exposition, we separate the gradient terms in Eq.~(\ref{grad-terms-6}), which yields
\begin{align}
\label{grad-terms-7}
\tilde{{\cal G}} &=-\frac{1}{2}\sum_{i,j}^N\,\gamma_{ij}\,\int d\mbfr\,c_{i}(\mbfr)\,c_{j}(\mbfr) \nonumber\\
&+\frac{1}{2}\sum_{i}^N \kappa_{ii}\int d\mbfr\,\left|\nabla{c}_{i}(\mbfr)\right|^2 \\
&+\frac{1}{2}\sum_{i,j\ne i}^N \kappa_{ij}\int d\mbfr\,\nabla c_{i}(\mbfr)\cdot\nabla{c}_{j}(\mbfr).\nonumber
\end{align}
The second term of Eq.~(\ref{grad-terms-7}) gives gradients terms of the Cahn-Hilliard form, while the third line yields cross terms. In this work, for simplicity,
we set $\kappa_{ij}=0$ for $i \ne j$. We note that such cross terms can become important when studying certain phenomena and/or when higher-order alloying interactions are considered.

\subsection{Correlation Kernels Containing Linear Terms in $n$}

To demonstrate the long wavelength limit of terms linear in density in Eq.~(\ref{multi-full-energy}), we consider, as an example,  the term
\begin{align}
\label{density-terms-1}
{\cal H} \!=\! -\frac{1}{2}\sum_{i,j}^N \int \! d\mbfr\,c_{i}^{o}c_{j}(\mbfr)\,\int d\mbfr^{\prime}\, C_2^{ij}(|\mbfr-\mbfr^{\prime}|)n(\mbfr^{\prime}).
\end{align}
Substituting the Fourier series expansion of the correlation function, Taylor expanding the correlation as in Eq.~(\ref{grad-terms-3}) (retaining the lowest order term), and taking the inverse Fourier transform yields,
\begin{align}
\label{density-terms-2}
\tilde{{\cal H}} &= -\frac{1}{2}\sum_{i,j}^N \gamma_{ij}  \int d\mbfr\,c_{i}^{o}\,c_{j}(\mbfr)n(\mbfr)\nonumber\\
&+\frac{1}{2}\sum_{i,j}^N \kappa_{ij}\int d\mbfr\,c_{i}^{o}c_{j}(\mbfr)\left(-\nabla^2\right)n(\mbfr),
\end{align}
where $\gamma_{ij}$ and $\kappa_{ij}$ are defined by Eqs.~(\ref{gamma}) and (\ref{kappa}), respectively.

The density, $n(\mbfr)$, in Eq.~(\ref{density-terms-2}) is rapidly varying. Its leading order representation is defined by a single-mode approximation of the form
\begin{align}
n(\mbfr)&= \sum_{m} A_{m}(\mbfr)\,e^{\mbfi\mbfq_{m}\cdot\mbfr},
\label{single-mode}
\end{align}
where $\mbfq_{m}$ are the reciprocal lattice vectors and $A_{m}(\mbfr)$ are slowly varying amplitudes corresponding to each reciprocal lattice vector, $m$. Substituting Eq.~(\ref{single-mode}) into Eq.~(\ref{density-terms-2}) gives
\begin{align}
\label{density-terms-3}
\tilde{{\cal H}} &=-\frac{1}{2}\sum_{i,j}^N\gamma_{i,j}\sum_{m} \int d\mbfr\,c_{i}^{o}\,c_{j}(\mbfr)A_{m}(\mbfr)\,e^{\mbfi\mbfq_{m}\cdot\mbfr}\nonumber\\
&-\frac{1}{2}\sum_{i,j}^N \kappa_{i,j}\sum_{m}\int d\mbfr\,c_{i}^{o}c_{j}(\mbfr)\nabla^2 \left( A_{m}(\mbfr)\,e^{\mbfi\mbfq_{m}\cdot\mbfr} \right).
\end{align}
Expanding the Laplacian in Eq.~(\ref{density-terms-3}) gives,
\begin{align}
\label{density-terms-4}
\tilde{{\cal H}} &=-\frac{1}{2}\sum_{i,j}^N\gamma_{i,j}\sum_{m} \int d\mbfr\,c_{i}^{o}\,c_{j}(\mbfr)A_{m}(\mbfr)\,e^{\mbfi\mbfq_{m}\cdot\mbfr}\nonumber\\
&-\frac{1}{2}\sum_{i,j}^N \kappa_{i,j}\sum_{m}\int d\mbfr\,c_{i}^{o}c_{j}(\mbfr)\,e^{\mbfi\mbfq_{m}\cdot\mbfr}{\cal L}_{m} A_{m}(\mbfr),
\end{align}
where ${\cal L}_{m}\equiv\nabla^2 +2\mbfi\mbfq_{m}\cdot\nabla-\mbfq_{m}^2$ is a covariant operator that assures rotational
invariance of the free energy in the long wavelength limit. It is noted that each term in Eq.~(\ref{density-terms-4}) only contains one rapidly oscillating variable, i.e., $e^{\mbfi\mbfq_{m}\cdot\mbfr}$. If we apply the so-called ``quick and dirty'' \cite{Goldenfeld05} analogue of the volume averaging method employed in
Refs.~\cite{Majaniemi09,Provatas10} (which amounts to decoupling slowly varying fields inside integrals from rapidly varying phase factors, thus
making the integrals effectively vanish when integrated over one unit), we obtain $\tilde{{\cal H}}\approx 0$.

It is straightforward to show that all other terms in Eq.~(\ref{multi-full-energy}) that are linear in $n$, such as,
\begin{align}
\label{linear-density+all}
{\cal H} =-\frac{1}{2}\sum_{i,j}^N \int d\mbfr\, n(\mbfr) c_{i}(\mbfr)\,\int d\mbfr^{\prime}\, C_2^{ij}(|\mbfr-\mbfr^{\prime}|)c_j(\mbfr^{\prime}),
\end{align}
similarly vanish upon coarse graining. It should also be evident from the above considerations that if  Eq.~(\ref{density-terms-1}) contained an $n(\mbfr) \cdots n(\mbfr')$ combination, then Eq.~(\ref{density-terms-4}) would contain terms with phase factors of different combinations of sums of two reciprocal lattice vectors.  Some of these two-vector combinations would add up to zero causing their corresponding terms to survive upon integration.

\subsection{Volume Averaging}
Equation~(\ref{density-terms-4}) can more formally be analyzed using a volume averaging convolution operator \cite{Majaniemi09},
defined by
\begin{align}
\label{volume-avg-1}
\langle  f(\mbfr) \rangle_V \equiv \frac{1}{\sqrt{\pi}V} \int_{-\infty}^{\infty} d\mbfr^{\prime} f(\mbfr') \chi_V(\mbfr-\mbfr'),
\end{align}
where $f(\mbfr^{\prime})$ is the function being course grained and $V$ is the coarse graining volume. The function $\chi_V$ in the integrand of Eq.~(\ref{volume-avg-1}) is a smoothing function that is normalized to unity, i.e.,
\begin{align}
\label{volume-avg-3}
\int_{-\infty}^{\infty} d\mbfr\,\chi_{V}(\mbfr-\mbfr^{\prime})\equiv 1.
\end{align}
A convenient form of  $\chi_V$ is given by
\begin{align}
\label{volume-avg-2}
\chi_V(\mbfr-\mbfr')  =  \frac{1}{\sqrt{\pi}V} \,e^{\frac{(\mbfr-\mbfr^{\prime})^2}{V^2}}.
\end{align}
In the long wavelength limit, $L_c \gg L \gg a$ where $L~\sim V^{1/d}$, in $d$-dimensions, while $L_c$ is the length scale of variation of the concentration field.
This condition implies that the function $\chi_V(\mbfr)$ varies on dimensions much larger than the lattice constant $a = 2\pi/|\mbfq_m|$ but much less then the length scale of variation of the concentration, $c_i(\mbfr)$.  Equation~(\ref{volume-avg-1}) defines a noninvertible limiting procedure that can be used to average a function over some volume.

It is instructive to apply the volume averaging procedure to the first term in Eq.~(\ref{density-terms-4}). For convenience we define $\phi(\mbfr) \equiv c_i^o c_j(\mbfr) A_m(\mbfr)$. It is noted that $\phi(\mbfr)$ varies on scales much larger than the lattice constant since it is comprised of slowly varying functions. Using the definition of $\phi(\mbfr)$, the first integral of Eq.~(\ref{density-terms-4}) can be written as
\begin{align}
&\tilde{{\cal H}}_V \!\! = \!\!
-\frac{1}{2}\sum_{i,j}^N\gamma_{i,j}\sum_{m} \! \int d\mbfr^{\prime} \!\! \left(\int \! d\mbfr \,\chi_V(\mbfr-\mbfr^{\prime})\right) \!\! \phi(\mbfr^{\prime}) e^{\mbfi\mbfq_{m}\cdot\mbfr^{\prime}}\nonumber\\
&=
-\frac{1}{2}\sum_{i,j}^N\gamma_{i,j}\sum_{m} \int d\mbfr \left(\int d\mbfr^{\prime}\, \chi_V(\mbfr-\mbfr^{\prime})\phi(\mbfr^{\prime})\,e^{\mbfi\mbfq_{m}\cdot\mbfr^{\prime}}\right),
\nonumber \\
\label{density-terms-5}
\end{align}
Since $\phi(\mbfr')$ varies more slowly than the scale of variation of $\chi_V$, it is reasonable to expand it in a Taylor series about $\mbfr'=\mbfr$. Substituting
$\phi(\mbfr')=\phi(\mbfr) - \nabla \phi(\mbfr) \cdot (\mbfr-\mbfr')$ into the above expression leads to
\begin{align}
&\tilde{{\cal H}}_V \!\! = \!\!
=-\frac{1}{2}\sum_{i,j}^N\gamma_{i,j}\sum_{m} \int d\mbfr  \bigg( \phi(\mbfr) \!\! \int \! \! d\mbfr^{\prime}\, \chi_V(\mbfr-\mbfr^{\prime})\,e^{\mbfi\mbfq_{m}\cdot\mbfr^{\prime}}\nonumber \\
& \!-\!  \nabla \phi(\mbfr) \! \cdot \! \int d \mbfr' \, \chi_V(\mbfr-\mbfr') \, (\mbfr-\mbfr^{\prime}) \, e^{\mbfi\mbfq_{m}\cdot\mbfr^{\prime} } \!+\! \cdots \bigg). \label{density-terms-6}
\end{align}
The noninvertible procedure was introduced in the second line of Eq.~(\ref{density-terms-5}). In the long wavelength limit, when $|\mbfq_m| L \rightarrow \infty$, both integrals in Eq.~(\ref{density-terms-6}) vanish as $\sim \left( |\mbfq_m L|\right)^{-1}$, making $\tilde{\cal H}_V$ similarly vanish.

\section{Phase Diagram Calculation}
\label{phase-diag-calc}
Equation~(\ref{coex}) provides a system of equations that are exact when one needs to determine the equilibrium properties of a given system. For a binary system, where the number of equations in Eq.~(\ref{coex}) is reduced by one, at a specified temperature and pressure they are sufficient to specify exactly the unique phase concentrations corresponding to the tie line between two phases.  However for multi-component systems, for the present case of a ternary alloy (represented by solute compositions $A$ and $B$), the set of conditions in Eq.~(\ref{coex}) are under-determined and cannot uniquely define all phase concentrations. This is because, for a ternary system at a specified temperature and pressure, there is not generally a single tie line which specifies phase boundaries between coexisting phases but, rather multiple tie lines defining the boundary between any two phases.

The under-determined set of conditions in Eq.~(\ref{coex}) contain variables $c_{A}^{I},c_{A}^{J},c_{B}^{I}$ and $c_{B}^{J}$ in phases $I$ and $J$ respectively.
To close this system, an additional condition is necessary to provide a fourth equation relating the concentrations. A convenient fourth condition is the {\it lever rule}, which relates weight fractions of phases to the average concentration. For clarity, we specify it here for ternary solid ($\alpha$) and liquid ($L$) phases,
\beq
\label{lever-A}
\bar{c}_{A} = c_{A}^{L}x_{L}+c_{A}^{\alpha}x_{\alpha}
\eeq
and
\beq
\label{lever-B}
\bar{c}_{B} = c_{B}^{L}x_{L}+c_{B}^{\alpha}x_{\alpha},
\eeq
where $\bar{c}_{A}$ and $\bar{c}_{B}$ are the average alloy compositions for components $A$ and $B$ respectively and $x_{L}$ and $x_{\alpha}$ represent the equilibrium volume fractions of liquid and $\alpha$ respectively, and satisfy $x_{L}+x_{\alpha}\equiv 1$. Combining this last relation between the volume fractions and Eqs.~(\ref{lever-A}) and (\ref{lever-B}) gives, the last equilibrium condition,
\beq
\label{lever-coex}
\frac{\bar{c}_{A}-c_{A}^{\alpha}}{c_{A}^{L}-c_{A}^{\alpha}}=\frac{\bar{c}_{B}-c_{B}^{\alpha}}{c_{B}^{L}-c_{B}^{\alpha}}.
\eeq
Equation~(\ref{coex}) together with Eq.~(\ref{lever-coex}) comprises a complete set of equations which can admit unique tie line solutions,
i.e., solutions for $c_{A}^{L},c_{A}^{\alpha},c_{B}^{L}$ and $c_{B}^{\alpha}$ in the solid-liquid example just considered.

With the free energy functions generally being highly nonlinear, it is not possible to find analytical solutions to Eqs.~(\ref{coex}) and (\ref{lever-coex}), and they must be solved numerically. One approach is to specify the temperature and then raster through the phase space of average  concentrations $\bar{c}_{A}$ and $\bar{c}_{B}$, where the rastering is done by taking discrete steps in steps of $\Delta\,c_A$ and $\Delta\,c_B$, respectively (for practical purposes its convenient to set $\Delta\,c_A=\Delta\,c_B=\Delta\,c$). For each pair of $\bar{c}_{A}$ and $\bar{c}_{B}$, Eqs.~(\ref{coex}) and (\ref{lever-coex}) can be solved numerically. The solutions yield $c_{A}^{L},c_{A}^{\alpha},c_{B}^{L}$ and $c_{B}^{\alpha}$. A unique solution for each pair of $\bar{c}_{A}$ and $\bar{c}_{B}$ defines one tie line. The collection of all such tie lines maps out the coexistence phase boundaries between any two phases, in the case considered here, $L$ and $\alpha$.
Where no solutions are admitted correspond to single phase regions where no tie lines exist. It is expected that the smoothness of the phase boundaries, when plotted for graphical purposes, will depend on the step size, $\Delta\,c$, chosen to discretize the average concentration values.

The above mentioned recipe can still require intensive computation, requiring a solution of four equations in four unknowns for $M^2$ combinations of average concentration pairs (where $M$ is the discretized number of average concentration values for a given component). Since this paper is intended to demonstrate the main features of our new multi-component (demonstrated for a ternary) PFC model, we adopted a simpler approach to compute the phase diagrams in section~\ref{ternary-eqm-properties}.  In particular, we fixed one of the equilibrium concentrations in Eq.~(\ref{coex}), assuming it is a valid solution at that temperature. We then solved for the remaining three unknown concentrations using Eq.~(\ref{coex}), repeating this $M$ times, once for each discrete value of the selected equilibrium concentration. Fixed concentrations were rastered in steps of $\Delta\,c$. Once again, a unique solution defines a tie line between coexisting phases, say $L$ and $\alpha$. If no solutions exist, we are in single phase regions where no tie lines exist.


\begin{thebibliography}{46}
\expandafter\ifx\csname natexlab\endcsname\relax\def\natexlab#1{#1}\fi
\expandafter\ifx\csname bibnamefont\endcsname\relax
  \def\bibnamefont#1{#1}\fi
\expandafter\ifx\csname bibfnamefont\endcsname\relax
  \def\bibfnamefont#1{#1}\fi
\expandafter\ifx\csname citenamefont\endcsname\relax
  \def\citenamefont#1{#1}\fi
\expandafter\ifx\csname url\endcsname\relax
  \def\url#1{\texttt{#1}}\fi
\expandafter\ifx\csname urlprefix\endcsname\relax\def\urlprefix{URL }\fi
\providecommand{\bibinfo}[2]{#2}
\providecommand{\eprint}[2][]{\url{#2}}

\bibitem[{\citenamefont{Karma and Rappel}(1998)}]{Karma98}
\bibinfo{author}{\bibfnamefont{A.}~\bibnamefont{Karma}} \bibnamefont{and}
  \bibinfo{author}{\bibfnamefont{W.~J.} \bibnamefont{Rappel}},
  \bibinfo{journal}{Phys. Rev. E} \textbf{\bibinfo{volume}{57}},
  \bibinfo{pages}{4323} (\bibinfo{year}{1998}).

\bibitem[{\citenamefont{Provatas et~al.}(1998)\citenamefont{Provatas,
  Goldenfeld, and Dantzig}}]{Provatas98}
\bibinfo{author}{\bibfnamefont{N.}~\bibnamefont{Provatas}},
  \bibinfo{author}{\bibfnamefont{N.}~\bibnamefont{Goldenfeld}},
  \bibnamefont{and} \bibinfo{author}{\bibfnamefont{J.}~\bibnamefont{Dantzig}},
  \bibinfo{journal}{Phys. Rev. Lett.} \textbf{\bibinfo{volume}{80}},
  \bibinfo{pages}{3308} (\bibinfo{year}{1998}).

\bibitem[{\citenamefont{Echebarria et~al.}(2004)\citenamefont{Echebarria,
  Folch, Karma, and Plapp}}]{Echebarria04}
\bibinfo{author}{\bibfnamefont{B.}~\bibnamefont{Echebarria}},
  \bibinfo{author}{\bibfnamefont{R.}~\bibnamefont{Folch}},
  \bibinfo{author}{\bibfnamefont{A.}~\bibnamefont{Karma}}, \bibnamefont{and}
  \bibinfo{author}{\bibfnamefont{M.}~\bibnamefont{Plapp}},
  \bibinfo{journal}{Phys. Rev. E.} \textbf{\bibinfo{volume}{70}},
  \bibinfo{pages}{061604} (\bibinfo{year}{2004}).

\bibitem[{\citenamefont{Greenwood et~al.}(2004)\citenamefont{Greenwood,
  Haataja, and Provatas}}]{Greenwood04}
\bibinfo{author}{\bibfnamefont{M.}~\bibnamefont{Greenwood}},
  \bibinfo{author}{\bibfnamefont{M.}~\bibnamefont{Haataja}}, \bibnamefont{and}
  \bibinfo{author}{\bibfnamefont{N.}~\bibnamefont{Provatas}},
  \bibinfo{journal}{Phys. Rev. Lett.} \textbf{\bibinfo{volume}{93}},
  \bibinfo{pages}{246101} (\bibinfo{year}{2004}).

\bibitem[{\citenamefont{Rappaz et~al.}(2003)\citenamefont{Rappaz, Jacot, and
  Boettinger}}]{Rappaz03}
\bibinfo{author}{\bibfnamefont{M.}~\bibnamefont{Rappaz}},
  \bibinfo{author}{\bibfnamefont{A.}~\bibnamefont{Jacot}}, \bibnamefont{and}
  \bibinfo{author}{\bibfnamefont{W.~J.} \bibnamefont{Boettinger}},
  \bibinfo{journal}{Metallurgical and Materials Transactions A}
  \textbf{\bibinfo{volume}{34}}, \bibinfo{pages}{467} (\bibinfo{year}{2003}).

\bibitem[{\citenamefont{Boettinger and Warren}(1999)}]{Boettinger99}
\bibinfo{author}{\bibfnamefont{W.~J.} \bibnamefont{Boettinger}}
  \bibnamefont{and} \bibinfo{author}{\bibfnamefont{J.~A.}
  \bibnamefont{Warren}}, \bibinfo{journal}{J. Crystal Growth}
  \textbf{\bibinfo{volume}{200}}, \bibinfo{pages}{583} (\bibinfo{year}{1999}).

\bibitem[{\citenamefont{Gr{\'a}n{\'a}sy
  et~al.}(2003)\citenamefont{Gr{\'a}n{\'a}sy, Pusztai, Warren, Douglas,
  B{\"o}rzs{\"o}nyi, and Ferreiro}}]{Granasy03}
\bibinfo{author}{\bibfnamefont{L.}~\bibnamefont{Gr{\'a}n{\'a}sy}},
  \bibinfo{author}{\bibfnamefont{T.}~\bibnamefont{Pusztai}},
  \bibinfo{author}{\bibfnamefont{J.~A.} \bibnamefont{Warren}},
  \bibinfo{author}{\bibfnamefont{J.~F.} \bibnamefont{Douglas}},
  \bibinfo{author}{\bibfnamefont{T.}~\bibnamefont{B{\"o}rzs{\"o}nyi}},
  \bibnamefont{and} \bibinfo{author}{\bibfnamefont{V.}~\bibnamefont{Ferreiro}},
  \bibinfo{journal}{Nature of Materials} \textbf{\bibinfo{volume}{{\bf 2}}},
  \bibinfo{pages}{92} (\bibinfo{year}{2003}).

\bibitem[{\citenamefont{Warren et~al.}(2000)\citenamefont{Warren, Kobayashi,
  and Carter}}]{Warran00}
\bibinfo{author}{\bibfnamefont{J.~A.} \bibnamefont{Warren}},
  \bibinfo{author}{\bibfnamefont{R.}~\bibnamefont{Kobayashi}},
  \bibnamefont{and} \bibinfo{author}{\bibfnamefont{W.~C.}
  \bibnamefont{Carter}}, \bibinfo{journal}{J. Cryst. Growth}
  \textbf{\bibinfo{volume}{{\bf 211}}}, \bibinfo{pages}{18}
  (\bibinfo{year}{2000}).

\bibitem[{\citenamefont{Warren et~al.}(2003)\citenamefont{Warren, Kobayashi,
  Lobkovsky, and Carter}}]{Warren03}
\bibinfo{author}{\bibfnamefont{J.~A.} \bibnamefont{Warren}},
  \bibinfo{author}{\bibfnamefont{R.}~\bibnamefont{Kobayashi}},
  \bibinfo{author}{\bibfnamefont{A.~E.} \bibnamefont{Lobkovsky}},
  \bibnamefont{and} \bibinfo{author}{\bibfnamefont{W.~C.}
  \bibnamefont{Carter}}, \bibinfo{journal}{Acta Materialia}
  \textbf{\bibinfo{volume}{{\bf 51}}}, \bibinfo{pages}{6035}
  (\bibinfo{year}{2003}).

\bibitem[{\citenamefont{Gr{\'a}n{\'a}sy
  et~al.}(2004)\citenamefont{Gr{\'a}n{\'a}sy, Pusztai, B{\"o}rzs{\"o}nyi,
  Warren, Kvamme, and James}}]{Granasy04a}
\bibinfo{author}{\bibfnamefont{L.}~\bibnamefont{Gr{\'a}n{\'a}sy}},
  \bibinfo{author}{\bibfnamefont{T.}~\bibnamefont{Pusztai}},
  \bibinfo{author}{\bibfnamefont{T.}~\bibnamefont{B{\"o}rzs{\"o}nyi}},
  \bibinfo{author}{\bibfnamefont{J.~A.} \bibnamefont{Warren}},
  \bibinfo{author}{\bibfnamefont{B.}~\bibnamefont{Kvamme}}, \bibnamefont{and}
  \bibinfo{author}{\bibfnamefont{P.~F.} \bibnamefont{James}},
  \bibinfo{journal}{Phys. Chem. Glasses} \textbf{\bibinfo{volume}{{\bf 45}}},
  \bibinfo{pages}{107} (\bibinfo{year}{2004}).

\bibitem[{\citenamefont{Bottger and Steinbach}(2006)}]{Steinbach06}
\bibinfo{author}{\bibfnamefont{B.}~\bibnamefont{Bottger}} \bibnamefont{and}
  \bibinfo{author}{\bibfnamefont{I.}~\bibnamefont{Steinbach}},
  \bibinfo{journal}{Acta Materialia} \textbf{\bibinfo{volume}{54}},
  \bibinfo{pages}{2697} (\bibinfo{year}{2006}).

\bibitem[{\citenamefont{Steinbach}(2009)}]{Steinbach09}
\bibinfo{author}{\bibfnamefont{I.}~\bibnamefont{Steinbach}},
  \bibinfo{journal}{Modelling Simul. Mater. Sci. Eng.}
  \textbf{\bibinfo{volume}{17}}, \bibinfo{pages}{073001}
  (\bibinfo{year}{2009}).

\bibitem[{\citenamefont{Choudhury and Nestler}(2012)}]{Nestler12}
\bibinfo{author}{\bibfnamefont{A.}~\bibnamefont{Choudhury}} \bibnamefont{and}
  \bibinfo{author}{\bibfnamefont{B.}~\bibnamefont{Nestler}},
  \bibinfo{journal}{Phys. Rev. E} \textbf{\bibinfo{volume}{85}},
  \bibinfo{pages}{021602} (\bibinfo{year}{2012}).

\bibitem[{\citenamefont{M.Haataja and L\'{e}onard}(2004)}]{Haataja04}
\bibinfo{author}{\bibnamefont{M.Haataja}} \bibnamefont{and}
  \bibinfo{author}{\bibfnamefont{F.}~\bibnamefont{L\'{e}onard}},
  \bibinfo{journal}{Phys. Rev. B} \textbf{\bibinfo{volume}{{\bf 69}}},
  \bibinfo{pages}{081201} (\bibinfo{year}{2004}).

\bibitem[{\citenamefont{Zhu et~al.}(2004)\citenamefont{Zhu, Wang, Ardell, Zhou,
  Lui, and Chen}}]{Zhu04a}
\bibinfo{author}{\bibfnamefont{J.~Z.} \bibnamefont{Zhu}},
  \bibinfo{author}{\bibfnamefont{T.}~\bibnamefont{Wang}},
  \bibinfo{author}{\bibfnamefont{A.~J.} \bibnamefont{Ardell}},
  \bibinfo{author}{\bibfnamefont{S.~H.} \bibnamefont{Zhou}},
  \bibinfo{author}{\bibfnamefont{Z.~K.} \bibnamefont{Lui}}, \bibnamefont{and}
  \bibinfo{author}{\bibfnamefont{L.~Q.} \bibnamefont{Chen}},
  \bibinfo{journal}{Acta Materialia} \textbf{\bibinfo{volume}{{\bf 52}}},
  \bibinfo{pages}{2837} (\bibinfo{year}{2004}).

\bibitem[{\citenamefont{Fan and Chen}(1996)}]{Fan96}
\bibinfo{author}{\bibfnamefont{D.}~\bibnamefont{Fan}} \bibnamefont{and}
  \bibinfo{author}{\bibfnamefont{L.-Q.} \bibnamefont{Chen}},
  \bibinfo{journal}{Acta Metallurgica} \textbf{\bibinfo{volume}{{\bf 45}}},
  \bibinfo{pages}{611} (\bibinfo{year}{1996}).

\bibitem[{\citenamefont{Wang et~al.}(2001)\citenamefont{Wang, Jin, Cuiti{\~n}o,
  and Khachaturyan}}]{Wang01}
\bibinfo{author}{\bibfnamefont{Y.}~\bibnamefont{Wang}},
  \bibinfo{author}{\bibfnamefont{Y.}~\bibnamefont{Jin}},
  \bibinfo{author}{\bibfnamefont{A.}~\bibnamefont{Cuiti{\~n}o}},
  \bibnamefont{and}
  \bibinfo{author}{\bibfnamefont{A.}~\bibnamefont{Khachaturyan}},
  \bibinfo{journal}{Acta Materialia} \textbf{\bibinfo{volume}{49}},
  \bibinfo{pages}{1847 } (\bibinfo{year}{2001}).

\bibitem[{\citenamefont{Elder et~al.}(2002)\citenamefont{Elder, Katakowski,
  Haataja, and Grant}}]{Elder02}
\bibinfo{author}{\bibfnamefont{K.~R.} \bibnamefont{Elder}},
  \bibinfo{author}{\bibfnamefont{M.}~\bibnamefont{Katakowski}},
  \bibinfo{author}{\bibfnamefont{M.}~\bibnamefont{Haataja}}, \bibnamefont{and}
  \bibinfo{author}{\bibfnamefont{M.}~\bibnamefont{Grant}},
  \bibinfo{journal}{Phys. Rev. Lett.} \textbf{\bibinfo{volume}{{\bf 88}}},
  \bibinfo{pages}{245701} (\bibinfo{year}{2002}).

\bibitem[{\citenamefont{Elder et~al.}(2007)\citenamefont{Elder, Provatas,
  Berry, Stefanovic, and Grant}}]{Elder07}
\bibinfo{author}{\bibfnamefont{K.~R.} \bibnamefont{Elder}},
  \bibinfo{author}{\bibfnamefont{N.}~\bibnamefont{Provatas}},
  \bibinfo{author}{\bibfnamefont{J.}~\bibnamefont{Berry}},
  \bibinfo{author}{\bibfnamefont{P.}~\bibnamefont{Stefanovic}},
  \bibnamefont{and} \bibinfo{author}{\bibfnamefont{M.}~\bibnamefont{Grant}},
  \bibinfo{journal}{Phys. Rev. B.} \textbf{\bibinfo{volume}{{\bf 75}}},
  \bibinfo{pages}{064107} (\bibinfo{year}{2007}).

\bibitem[{\citenamefont{Jin and Khachaturyan}(2006)}]{Jin06}
\bibinfo{author}{\bibfnamefont{Y.~M.} \bibnamefont{Jin}} \bibnamefont{and}
  \bibinfo{author}{\bibfnamefont{A.~G.} \bibnamefont{Khachaturyan}},
  \bibinfo{journal}{Journal of Applied Physics} \textbf{\bibinfo{volume}{100}},
  \bibinfo{pages}{013519} (\bibinfo{year}{2006}).

\bibitem[{\citenamefont{Jaatinen et~al.}(2009)\citenamefont{Jaatinen, Achim,
  Elder, and Ala-Nissila}}]{Akusti09}
\bibinfo{author}{\bibfnamefont{A.}~\bibnamefont{Jaatinen}},
  \bibinfo{author}{\bibfnamefont{C.~V.} \bibnamefont{Achim}},
  \bibinfo{author}{\bibfnamefont{K.~R.} \bibnamefont{Elder}}, \bibnamefont{and}
  \bibinfo{author}{\bibfnamefont{T.}~\bibnamefont{Ala-Nissila}},
  \bibinfo{journal}{Phys. Rev. E} \textbf{\bibinfo{volume}{{\bf 80}}},
  \bibinfo{pages}{031602} (\bibinfo{year}{2009}).

\bibitem[{\citenamefont{Berry et~al.}(2008)\citenamefont{Berry, Elder, and
  Grant}}]{Berry08}
\bibinfo{author}{\bibfnamefont{J.}~\bibnamefont{Berry}},
  \bibinfo{author}{\bibfnamefont{K.~R.} \bibnamefont{Elder}}, \bibnamefont{and}
  \bibinfo{author}{\bibfnamefont{M.}~\bibnamefont{Grant}},
  \bibinfo{journal}{Phys. Rev. B} \textbf{\bibinfo{volume}{77}},
  \bibinfo{pages}{224114} (\bibinfo{year}{2008}).

\bibitem[{\citenamefont{Archer et~al.}(2012)\citenamefont{Archer, Robbins, and
  Thiele}}]{Archer12}
\bibinfo{author}{\bibfnamefont{A.~J.} \bibnamefont{Archer}},
  \bibinfo{author}{\bibfnamefont{M.~J.} \bibnamefont{Robbins}},
  \bibnamefont{and} \bibinfo{author}{\bibfnamefont{U.}~\bibnamefont{Thiele}},
  \bibinfo{journal}{Phys. Rev. E} \textbf{\bibinfo{volume}{{\bf 86}}},
  \bibinfo{pages}{031603} (\bibinfo{year}{2012}).

\bibitem[{\citenamefont{Huang and Elder}(2008)}]{Huang08}
\bibinfo{author}{\bibfnamefont{Z.-F.} \bibnamefont{Huang}} \bibnamefont{and}
  \bibinfo{author}{\bibfnamefont{K.~R.} \bibnamefont{Elder}},
  \bibinfo{journal}{Phys. Rev. Lett.} \textbf{\bibinfo{volume}{101}},
  \bibinfo{pages}{158701} (\bibinfo{year}{2008}).

\bibitem[{\citenamefont{Wu and Karma}(2007)}]{Wu07}
\bibinfo{author}{\bibfnamefont{K.-A.} \bibnamefont{Wu}} \bibnamefont{and}
  \bibinfo{author}{\bibfnamefont{A.}~\bibnamefont{Karma}},
  \bibinfo{journal}{Phys. Rev. B} \textbf{\bibinfo{volume}{76}},
  \bibinfo{pages}{184107} (\bibinfo{year}{2007}).

\bibitem[{\citenamefont{Majaniemi and Provatas}(2009)}]{Majaniemi09}
\bibinfo{author}{\bibfnamefont{S.}~\bibnamefont{Majaniemi}} \bibnamefont{and}
  \bibinfo{author}{\bibfnamefont{N.}~\bibnamefont{Provatas}},
  \bibinfo{journal}{Phys. Rev. E} \textbf{\bibinfo{volume}{79}},
  \bibinfo{pages}{011607} (\bibinfo{year}{2009}).

\bibitem[{\citenamefont{Provatas and Majaniemi}(2010)}]{Provatas10}
\bibinfo{author}{\bibfnamefont{N.}~\bibnamefont{Provatas}} \bibnamefont{and}
  \bibinfo{author}{\bibfnamefont{S.}~\bibnamefont{Majaniemi}},
  \bibinfo{journal}{Phys. Rev. E} \textbf{\bibinfo{volume}{82}},
  \bibinfo{pages}{041601} (\bibinfo{year}{2010}).

\bibitem[{\citenamefont{Athreya et~al.}(2007)\citenamefont{Athreya, Goldenfeld,
  Dantzig, Greenwood, and Provatas}}]{Athreya07}
\bibinfo{author}{\bibfnamefont{B.~P.} \bibnamefont{Athreya}},
  \bibinfo{author}{\bibfnamefont{N.}~\bibnamefont{Goldenfeld}},
  \bibinfo{author}{\bibfnamefont{J.~A.} \bibnamefont{Dantzig}},
  \bibinfo{author}{\bibfnamefont{M.}~\bibnamefont{Greenwood}},
  \bibnamefont{and} \bibinfo{author}{\bibfnamefont{N.}~\bibnamefont{Provatas}},
  \bibinfo{journal}{Phys. Rev. E} \textbf{\bibinfo{volume}{76}},
  \bibinfo{pages}{056706} (\bibinfo{year}{2007}).

\bibitem[{\citenamefont{Greenwood et~al.}(2010)\citenamefont{Greenwood,
  Provatas, and Rottler}}]{Greenwood10}
\bibinfo{author}{\bibfnamefont{M.}~\bibnamefont{Greenwood}},
  \bibinfo{author}{\bibfnamefont{N.}~\bibnamefont{Provatas}}, \bibnamefont{and}
  \bibinfo{author}{\bibfnamefont{J.}~\bibnamefont{Rottler}},
  \bibinfo{journal}{Phys. Rev. Lett.} \textbf{\bibinfo{volume}{105}},
  \bibinfo{pages}{045702} (\bibinfo{year}{2010}).

\bibitem[{\citenamefont{Greenwood
  et~al.}(2011{\natexlab{a}})\citenamefont{Greenwood, Rottler, and
  Provatas}}]{Greenwood11}
\bibinfo{author}{\bibfnamefont{M.}~\bibnamefont{Greenwood}},
  \bibinfo{author}{\bibfnamefont{J.}~\bibnamefont{Rottler}}, \bibnamefont{and}
  \bibinfo{author}{\bibfnamefont{N.}~\bibnamefont{Provatas}},
  \bibinfo{journal}{Phys. Rev. E} \textbf{\bibinfo{volume}{83}},
  \bibinfo{pages}{031601} (\bibinfo{year}{2011}{\natexlab{a}}).

\bibitem[{\citenamefont{Greenwood
  et~al.}(2011{\natexlab{b}})\citenamefont{Greenwood, Ofori-Opoku, Rottler, and
  Provatas}}]{GreenwoodOfori11}
\bibinfo{author}{\bibfnamefont{M.}~\bibnamefont{Greenwood}},
  \bibinfo{author}{\bibfnamefont{N.}~\bibnamefont{Ofori-Opoku}},
  \bibinfo{author}{\bibfnamefont{J.}~\bibnamefont{Rottler}}, \bibnamefont{and}
  \bibinfo{author}{\bibfnamefont{N.}~\bibnamefont{Provatas}},
  \bibinfo{journal}{Phys. Rev. B} \textbf{\bibinfo{volume}{84}},
  \bibinfo{pages}{064104} (\bibinfo{year}{2011}{\natexlab{b}}).

\bibitem[{\citenamefont{Greenwood et~al.}(2012)\citenamefont{Greenwood,
  Sinclair, and Millitzer}}]{Greenwood12}
\bibinfo{author}{\bibfnamefont{M.}~\bibnamefont{Greenwood}},
  \bibinfo{author}{\bibfnamefont{C.}~\bibnamefont{Sinclair}}, \bibnamefont{and}
  \bibinfo{author}{\bibfnamefont{M.}~\bibnamefont{Millitzer}},
  \bibinfo{journal}{Acta. Materialia} \textbf{\bibinfo{volume}{60}},
  \bibinfo{pages}{5752} (\bibinfo{year}{2012}).

\bibitem[{\citenamefont{Rottler et~al.}(2012)\citenamefont{Rottler, Greenwood,
  and Ziebarth}}]{rottler12}
\bibinfo{author}{\bibfnamefont{J.}~\bibnamefont{Rottler}},
  \bibinfo{author}{\bibfnamefont{M.}~\bibnamefont{Greenwood}},
  \bibnamefont{and} \bibinfo{author}{\bibfnamefont{B.}~\bibnamefont{Ziebarth}},
  \bibinfo{journal}{J. Phys.: Condens. Matter} \textbf{\bibinfo{volume}{24}},
  \bibinfo{pages}{135002} (\bibinfo{year}{2012}).

\bibitem[{\citenamefont{Fallah et~al.}(2012{\natexlab{a}})\citenamefont{Fallah,
  Stolle, Ofori-Opoku, Esmaeili, and Provatas}}]{Fallah12a}
\bibinfo{author}{\bibfnamefont{V.}~\bibnamefont{Fallah}},
  \bibinfo{author}{\bibfnamefont{J.}~\bibnamefont{Stolle}},
  \bibinfo{author}{\bibfnamefont{N.}~\bibnamefont{Ofori-Opoku}},
  \bibinfo{author}{\bibfnamefont{S.}~\bibnamefont{Esmaeili}}, \bibnamefont{and}
  \bibinfo{author}{\bibfnamefont{N.}~\bibnamefont{Provatas}},
  \bibinfo{journal}{Phy. Rev. B} \textbf{\bibinfo{volume}{86}},
  \bibinfo{pages}{134112} (\bibinfo{year}{2012}{\natexlab{a}}).

\bibitem[{\citenamefont{Berry et~al.}(2012)\citenamefont{Berry, Provatas,
  Rottler, and Sinclair}}]{Berry12}
\bibinfo{author}{\bibfnamefont{J.}~\bibnamefont{Berry}},
  \bibinfo{author}{\bibfnamefont{N.}~\bibnamefont{Provatas}},
  \bibinfo{author}{\bibfnamefont{J.}~\bibnamefont{Rottler}}, \bibnamefont{and}
  \bibinfo{author}{\bibfnamefont{C.~W.} \bibnamefont{Sinclair}},
  \bibinfo{journal}{Submitted to Phys. Rev. B,}
  \textbf{\bibinfo{volume}{http://arxiv.org/abs/1210.1527}}
  (\bibinfo{year}{2012}).

\bibitem[{\citenamefont{Ramakrishan and Yussouff}(1979)}]{Ramakrishan79}
\bibinfo{author}{\bibfnamefont{T.~V.} \bibnamefont{Ramakrishan}}
  \bibnamefont{and} \bibinfo{author}{\bibfnamefont{M.}~\bibnamefont{Yussouff}},
  \bibinfo{journal}{Phy. Rev. B} \textbf{\bibinfo{volume}{19}},
  \bibinfo{pages}{2775} (\bibinfo{year}{1979}).

\bibitem[{\citenamefont{Elder et~al.}(2010)\citenamefont{Elder, Huang, and
  Provatas}}]{Elder10}
\bibinfo{author}{\bibfnamefont{K.~R.} \bibnamefont{Elder}},
  \bibinfo{author}{\bibfnamefont{Z.-F.} \bibnamefont{Huang}}, \bibnamefont{and}
  \bibinfo{author}{\bibfnamefont{N.}~\bibnamefont{Provatas}},
  \bibinfo{journal}{Phys. Rev. E} \textbf{\bibinfo{volume}{81}},
  \bibinfo{pages}{011602} (\bibinfo{year}{2010}).

\bibitem[{\citenamefont{Huang et~al.}(2010)\citenamefont{Huang, Elder, and
  Provatas}}]{Huang10}
\bibinfo{author}{\bibfnamefont{Z.-F.} \bibnamefont{Huang}},
  \bibinfo{author}{\bibfnamefont{K.~R.} \bibnamefont{Elder}}, \bibnamefont{and}
  \bibinfo{author}{\bibfnamefont{N.}~\bibnamefont{Provatas}},
  \bibinfo{journal}{Phys. Rev. E} \textbf{\bibinfo{volume}{82}},
  \bibinfo{pages}{021605} (\bibinfo{year}{2010}).

\bibitem[{\citenamefont{Tegze et~al.}(2011)\citenamefont{Tegze, T\'oth, and
  Gr\'an\'asy}}]{Tegze11}
\bibinfo{author}{\bibfnamefont{G.}~\bibnamefont{Tegze}},
  \bibinfo{author}{\bibfnamefont{G.~I.} \bibnamefont{T\'oth}},
  \bibnamefont{and}
  \bibinfo{author}{\bibfnamefont{L.}~\bibnamefont{Gr\'an\'asy}},
  \bibinfo{journal}{Phys. Rev. Lett.} \textbf{\bibinfo{volume}{106}},
  \bibinfo{pages}{195502} (\bibinfo{year}{2011}).

\bibitem[{\citenamefont{Fallah et~al.}(2012{\natexlab{b}})\citenamefont{Fallah,
  Ofori-Opoku, Stolle, Provatas, and Esmaeili}}]{Fallah12b}
\bibinfo{author}{\bibfnamefont{V.}~\bibnamefont{Fallah}},
  \bibinfo{author}{\bibfnamefont{N.}~\bibnamefont{Ofori-Opoku}},
  \bibinfo{author}{\bibfnamefont{J.}~\bibnamefont{Stolle}},
  \bibinfo{author}{\bibfnamefont{N.}~\bibnamefont{Provatas}}, \bibnamefont{and}
  \bibinfo{author}{\bibfnamefont{S.}~\bibnamefont{Esmaeili}},
  \bibinfo{journal}{Submitted to Acta materialia,}
  \textbf{\bibinfo{volume}{http://arxiv.org/abs/1210.4977}}
  (\bibinfo{year}{2012}{\natexlab{b}}).

\bibitem[{\citenamefont{Raghavan}(2007)}]{Raghavan07}
\bibinfo{author}{\bibfnamefont{V.}~\bibnamefont{Raghavan}},
  \bibinfo{journal}{Journal of Phase Equilibria and Diffusion}
  \textbf{\bibinfo{volume}{28}}, \bibinfo{pages}{174} (\bibinfo{year}{2007}).

\bibitem[{\citenamefont{Ozawa and Kimura}(1970)}]{Ozawa70}
\bibinfo{author}{\bibfnamefont{E.}~\bibnamefont{Ozawa}} \bibnamefont{and}
  \bibinfo{author}{\bibfnamefont{H.}~\bibnamefont{Kimura}},
  \bibinfo{journal}{Acta Metall.} \textbf{\bibinfo{volume}{18}},
  \bibinfo{pages}{995 } (\bibinfo{year}{1970}), ISSN \bibinfo{issn}{0001-6160}.

\bibitem[{\citenamefont{Somoza et~al.}(2002)\citenamefont{Somoza, Petkov, Lynn,
  and Dupasquier}}]{Somoza02}
\bibinfo{author}{\bibfnamefont{A.}~\bibnamefont{Somoza}},
  \bibinfo{author}{\bibfnamefont{M.~P.} \bibnamefont{Petkov}},
  \bibinfo{author}{\bibfnamefont{K.~G.} \bibnamefont{Lynn}}, \bibnamefont{and}
  \bibinfo{author}{\bibfnamefont{A.}~\bibnamefont{Dupasquier}},
  \bibinfo{journal}{Phys. Rev. B} \textbf{\bibinfo{volume}{65}},
  \bibinfo{pages}{094107} (\bibinfo{year}{2002}).

\bibitem[{\citenamefont{Nagai et~al.}(2001)\citenamefont{Nagai, Murayama, Tang,
  Nonaka, Hono, and Hasegawa}}]{Nagai01}
\bibinfo{author}{\bibfnamefont{Y.}~\bibnamefont{Nagai}},
  \bibinfo{author}{\bibfnamefont{M.}~\bibnamefont{Murayama}},
  \bibinfo{author}{\bibfnamefont{Z.}~\bibnamefont{Tang}},
  \bibinfo{author}{\bibfnamefont{T.}~\bibnamefont{Nonaka}},
  \bibinfo{author}{\bibfnamefont{K.}~\bibnamefont{Hono}}, \bibnamefont{and}
  \bibinfo{author}{\bibfnamefont{M.}~\bibnamefont{Hasegawa}},
  \bibinfo{journal}{Acta materialia} \textbf{\bibinfo{volume}{49}},
  \bibinfo{pages}{913} (\bibinfo{year}{2001}).

\bibitem[{\citenamefont{Babu et~al.}(2012)\citenamefont{Babu, Rajaraman,
  Amarendra, Govindaraj, Lalla, Dasgupta, Bhalerao, and Sundar}}]{Babu12}
\bibinfo{author}{\bibfnamefont{S.~H.} \bibnamefont{Babu}},
  \bibinfo{author}{\bibfnamefont{R.}~\bibnamefont{Rajaraman}},
  \bibinfo{author}{\bibfnamefont{G.}~\bibnamefont{Amarendra}},
  \bibinfo{author}{\bibfnamefont{R.}~\bibnamefont{Govindaraj}},
  \bibinfo{author}{\bibfnamefont{N.~P.} \bibnamefont{Lalla}},
  \bibinfo{author}{\bibfnamefont{A.}~\bibnamefont{Dasgupta}},
  \bibinfo{author}{\bibfnamefont{G.}~\bibnamefont{Bhalerao}}, \bibnamefont{and}
  \bibinfo{author}{\bibfnamefont{C.~S.} \bibnamefont{Sundar}},
  \bibinfo{journal}{Philosophical Magazine} \textbf{\bibinfo{volume}{92}},
  \bibinfo{pages}{2848} (\bibinfo{year}{2012}).

\bibitem[{\citenamefont{Goldenfeld et~al.}(2005)\citenamefont{Goldenfeld,
  Athreya, and Dantzig}}]{Goldenfeld05}
\bibinfo{author}{\bibfnamefont{N.}~\bibnamefont{Goldenfeld}},
  \bibinfo{author}{\bibfnamefont{B.~P.} \bibnamefont{Athreya}},
  \bibnamefont{and} \bibinfo{author}{\bibfnamefont{J.~A.}
  \bibnamefont{Dantzig}}, \bibinfo{journal}{Phys. Rev. E.}
  \textbf{\bibinfo{volume}{{\bf 72}}}, \bibinfo{pages}{020601}
  (\bibinfo{year}{2005}).

\end{thebibliography}

\end{document}